\documentclass[twocolumn,twocolappendix]{aastex631}

\usepackage{amsmath}
\usepackage{amssymb}
\usepackage{graphicx}
\graphicspath{{figures/}}
\usepackage{natbib}
\usepackage{booktabs}
\usepackage{multirow}
\usepackage{xspace}
\usepackage{placeins}  



\newcommand{\wise}{WISE\xspace}
\newcommand{\spitzer}{Spitzer\xspace}
\newcommand{\cosmos}{COSMOS\xspace}
\newcommand{\neowise}{NEOWISE\xspace}
\newcommand{\irac}{IRAC\xspace}
\newcommand{\unwise}{unWISE\xspace}
\newcommand{\MJysr}{MJy\,sr$^{-1}$\xspace}

\shorttitle{Super-Resolution: WISE to Spitzer}
\shortauthors{Rezaee et al.}

\begin{document}

\title{Enhancing WISE Infrared Imaging to Spitzer Resolution Using Deep Learning Super-Resolution}

\correspondingauthor{Saeed Rezaee}
\email{sreza003@ucr.edu}
\author[0000-0002-6289-9918]{Saeed Rezaee}
\affiliation{Department of Physics and Astronomy, University of California, Riverside, CA 92521, USA}

\author[0000-0003-2226-5395]{Shoubaneh Hemmati}
\affiliation{IPAC, California Institute of Technology, 1200 E. California Blvd. Pasadena, CA 91125, USA}

\author[0000-0001-5846-4404]{Bahram Mobasher}
\affiliation{Department of Physics and Astronomy, University of California, Riverside, CA 92521, USA}

\author[0000-0001-9687-4973]{Naveen A. Reddy}
\affiliation{Department of Physics and Astronomy, University of California, 900 University Ave, Riverside, CA 92521, USA}

\author[0000-0002-3551-279X]{Tara Fetherolf}
\affiliation{Department of Earth and Planetary Sciences, University of California Riverside, Riverside, California, USA; \dag NASA Postdoctoral}

\author[0000-0002-6219-5558]{Alexander de la Vega}
\affiliation{Department of Physics and Astronomy, University of California, 900 University Ave, Riverside, CA 92521, USA}

\author[0009-0001-7571-7562]{Negin Moharrami Allafi}
\affiliation{The Graduate Center, CUNY, New York, NY 10016, USA}

\author[0009-0006-6913-1561]{Sina Miri}
\affiliation{Department of Mechanical and Industrial Engineering, University of Illinois at Chicago, Chicago, IL 60607, USA}

\begin{abstract}
We present a deep-learning framework that performs $4.6\times$ spatial super-resolution from \wise W1 (3.4\,\micron) toward \spitzer \irac Ch1 (3.6\,\micron), and characterize its behavior on the \cosmos field. We train on $\sim 390{,}000$ paired cutouts drawn uniformly within the \wise/\spitzer overlap and report all metrics on a held-out test set of 83{,}592 cutouts. The framework uses a convolutional neural network trained with a loss function that emphasizes accurate recovery of sources in crowded fields. The model recovers the total flux of the central source in a fixed aperture to a median relative error of 11\%, a factor of $\approx 2$ better than the interpolation baselines; the gain reaches $\approx 3\times$ on the faintest quartile. The brightness dependence is monotonic: the aperture integrated flux error decreases from 13\% on the faintest quartile to 8\% on the brightest. At the 3--5\arcsec\ separations where \wise\ blends sources that \spitzer\ separates, the model recovers 35\% of the source peaks detectable in the Spitzer truth compared with 9\% for interpolation. The characteristic failure mode is oversmoothing of source profiles, which biases integrated flux measurements upward; this pattern is qualitatively similar to that of the interpolation baselines but is quantitatively smaller for the trained model. These results suggest genuine resolution enhancement and source deblending, providing a path toward applying super-resolution across the all-sky area that \spitzer could not cover. An appendix replicates the analysis at W2 $\to$ \irac Ch2 with consistent results; the trained model and code are publicly available.
\end{abstract}


\section{INTRODUCTION}
\label{sec:introduction}

Observations of infrared emission ($\approx 1\,\micron$--1\,mm) from galaxies probe processes that are difficult or impossible to access at optical wavelengths, including dust-obscured star formation \citep[e.g.,][]{Sanders1996, Calzetti2000, Chary2001, Casey2014, Traina2024}, interstellar dust emission \citep[e.g.,][]{Draine2007, Lagache2005}, the rest-frame near-infrared light of evolved stellar populations that traces galaxy stellar mass, and the rest-frame optical emission of galaxies at $z \gtrsim 1$ redshifted into the observed infrared \citep[e.g.,][]{Elbaz2011}. The science return from infrared imaging depends not only on sensitivity but on angular resolution: at fixed flux limit, the recovered source density, photometric accuracy, and morphological information all scale with the ability to deblend nearby sources at separations comparable to the instrumental point-spread function (PSF).

Image restoration has long been central to extracting science from astronomical images. Classical approaches include maximum-entropy deconvolution \citep{Narayan1986}, Wiener filtering \citep{Zaroubi1995}, Richardson-Lucy iteration \citep{Richardson1972, Lucy1974}, drizzling of dithered exposures \citep{FruchterHook2002}, lucky imaging \citep{Law2006}, and multiscale CLEAN-style deconvolution for interferometric data \citep[e.g.,][]{Rau2011, Offringa2017}. Together they have shaped how PSF blur, undersampling, and correlated noise are mitigated in surveys from radio through optical wavelengths. These methods are limited, however, by hand-tuned priors and by linearity assumptions that break down when the resolution gap between low- and high-fidelity data is large, motivating a shift toward learned image priors.

Convolutional super-resolution \citep{Dong2014} substantially extended what can be recovered below the nominal diffraction limit. Successive refinements built on that foundation: perceptual and adversarial losses \citep{Ledig2017}, very-deep residual scaling \citep{Lim2017EDSR}, efficient subpixel upsampling \citep{Shi2016}, squeeze-and-excitation channel reweighting \citep{Hu2018SENet}, residual channel-attention networks \citep{Zhang2018RCAN}, enhanced GAN-based reconstruction \citep{Wang2018ESRGAN}, and transformer-based restoration \citep{Liang2021SwinIR}. These architectures have been transferred to astronomical data with growing success, with \citet{Schawinski2017} recovering galaxy features beyond the deconvolution limit, and further applications spanning solar imaging \citep[e.g.,][]{Diaz2018, Song2024, Ramunno2025}, radio surveys \citep[e.g.,][]{Gheller2018, Dabbech2022}, X-ray data \citep{Sweere2022}, strong gravitational lensing \citep{ReddyP2024}, optical ground-based imaging \citep[e.g.,][]{Sukurdeep2025}, and galaxy morphology classification \citep[e.g.,][]{Dieleman2015, HuertasCompany2023}. Comparatively little work, however, has addressed the infrared regime, where the resolution gap between all-sky surveys such as \wise and pointed, diffraction-limited imaging from \spitzer or JWST is largest and where bridging it is therefore the most beneficial. The infrared lag has two practical causes. Wide-area infrared surveys use modest apertures at relatively long wavelengths, so their diffraction-limited point-spread functions are coarse, and although infrared interferometry can deliver long-baseline resolution, it is not generally available for wide-field survey imaging. Supervised reconstruction of the kind used here also benefits from co-registered low- and high-resolution imaging at closely matched wavelengths, and for this instrument pair such training data exist only in fields that the higher-resolution pointed facility has also observed.

The Wide-field Infrared Survey Explorer \citep[\wise{};][]{Wright2010} has surveyed the entire sky at 3.4, 4.6, 12, and 22\,\micron, with the \neowise reactivation \citep{Mainzer2014} extending the coverage from 2013 to the mission's decommissioning in 2024 and yielding more than a decade of time-domain data. The W1 band reaches a 5$\sigma$ point-source sensitivity of $\approx 54\,\mu$Jy in the AllWISE release \citep{Cutri2013}. The spatial resolution of \wise is, however, comparatively coarse: the W1 band (3.4\,\micron) has a PSF full-width at half-maximum (FWHM) of $\approx 6.1\arcsec$ at a native pixel sampling of 2.75\arcsec\ pixel$^{-1}$. This resolution is adequate for detection and cataloging, but limits studies of galaxy morphology, source deblending in crowded fields, and characterization of faint structures near bright objects.

The \spitzer Space Telescope Infrared Array Camera \citep[\irac{};][]{Fazio2004} reached substantially finer spatial resolution at similar wavelengths. The 3.6\,\micron\ channel (Ch1) achieved a PSF FWHM of $\approx 1.7\arcsec$ with a drizzled pixel scale of 0.6\arcsec\ pixel$^{-1}$ in both cryogenic and warm-mission phases, enabling galaxy structural decomposition, photometry in crowded regions, and identification of faint companions. \spitzer concluded operations in 2020, however, and its pointed observation strategy meant that only selected fields received deep \irac imaging.

The combination of \wise's all-sky coverage with \spitzer-class angular resolution is therefore the natural target of a super-resolution approach. The multi-epoch cadence with \wise\ enables time-domain science that would benefit from improved deblending and photometric accuracy, and the Roman and Euclid Space Telescopes are expected to benefit from enhanced infrared priors for source identification.

In this paper, we present and characterize a deep-learning framework that enhances \wise W1 images toward \spitzer \irac Ch1 resolution. Our approach uses the \cosmos field \citep{Scoville2007} as training data, where deep \spitzer \irac imaging and complete \wise coverage provide well-characterized paired observations. We adapt the Residual Channel Attention Network of \citet{Zhang2018RCAN}, which we call the Enhanced RCAN, for the low-resolution, source-crowded cutouts typical of deep \cosmos imaging (Section~\ref{sec:methods}). We then characterize the model beyond standard image quality metrics, measuring aperture photometry accuracy as a function of source brightness, source deblending recovery in the regime where \wise blends sources that \spitzer separates, and faint companion recovery against catalog ground truth.

The outline of this paper is as follows. In Section~\ref{sec:data}, we describe the \wise and \spitzer datasets and the \cosmos training field. In Section~\ref{sec:methods}, we detail the network architecture and training procedure. In Section~\ref{sec:results}, we characterize the model's behavior across image quality, photometric fidelity, and source recovery, including its dependence on source brightness, its performance on source deblending and faint companion detection, and its oversmoothing failure mode, and close with the domain of validity that the characterization defines. In Section~\ref{sec:discussion}, we interpret what the network has learned and relate the results to previous super-resolution efforts. We summarize the results in Section~\ref{sec:conclusions}. Appendix~\ref{sec:appendix_ch2} replicates the full analysis at the W2 $\to$ \irac Ch2 channel pair.

\section{DATA}
\label{sec:data}
Training our framework requires the same sky regions imaged by both \wise and \spitzer. We use the \cosmos field \citep{Scoville2007}, which has coverage from both missions.

\subsection{Wide-field Infrared Survey Explorer (WISE)}
\label{sec:wise}

\wise surveyed the entire sky in four infrared bands: W1 (3.4\,\micron), W2 (4.6\,\micron), W3 (12\,\micron), and W4 \citep[22\,\micron;][]{Wright2010}. The spacecraft carries a 40\,cm telescope feeding a 1024$\times$1024 pixel detector array per band. The W1 channel, on which we focus in the main text, has a native pixel scale of 2.75\arcsec\ and achieves a 5$\sigma$ point-source sensitivity of $\approx 54\,\mu$Jy in the AllWISE data release \citep{Cutri2013}. Appendix~\ref{sec:appendix_ch2} replicates the entire pipeline, training, and evaluation suite at W2 $\to$ \irac Ch2, with consistent results.

We adopt the \unwise image stacks \citep{Lang2014, Meisner2017, unWISEimages2021}, a standard community reprocessing of the \wise imaging, which coadd the \wise and \neowise imaging at the native \wise pixel scale without the PSF-convolution (blurring) step applied to the AllWISE Atlas Images, preserving the intrinsic $\approx 6\arcsec$ resolution, with deeper effective coverage from the full stack of epochs. The \unwise images are calibrated in linear Vega nanomaggies (nMgy), with a photometric zero point MAGZP\,=\,22.5 defining the corresponding Vega magnitudes.

The W1 PSF is approximately Gaussian with FWHM\,$\approx 6.1\arcsec$, though the actual PSF carries non-Gaussian wings and slight asymmetry from optical effects. At $z = 0.5$, this resolution corresponds to a physical scale of $\approx 37$\,kpc, larger than the stellar extent of a typical galaxy, sufficient to limit studies of galaxy substructure at cosmologically interesting distances.

We note that the \unwise stacks contain a substantial fraction ($\approx 43\%$ in our dataset) of negative pixel values arising from the sky-subtraction process; these represent genuine background fluctuations rather than missing data, and proper handling of these values is essential for accurate super-resolution, as discussed in Section~\ref{sec:preprocessing}.

\subsection{Spitzer Space Telescope}
\label{sec:spitzer}

The \spitzer Space Telescope operated from 2003 to 2020, providing infrared imaging and spectroscopy across 3.6--160\,\micron~\citep{Werner2004}. The Infrared Array Camera \citep[\irac{};][]{Fazio2004} imaged at 3.6, 4.5, 5.8, and 8.0\,\micron\ during cryogenic operations (2003--2009), with the 3.6 and 4.5\,\micron\ channels continuing through the warm mission (2009--2020).

The \irac 3.6\,\micron\ channel (Ch1) uses a 256$\times$256 pixel InSb detector array with a native pixel scale of 1.22\arcsec. Deep imaging programs typically dither and mosaic observations to a drizzled scale of 0.6\arcsec\ pixel$^{-1}$, which we adopt as our target pixel scale. The Ch1 PSF has FWHM\,$\approx 1.66\arcsec$, $\approx 3.7\times$ finer than \wise W1, which sets the target resolution.

\spitzer \irac mosaics are calibrated in surface-brightness units of \MJysr. The conversion from pixel flux density to surface brightness depends on the per-pixel solid angle:
\begin{equation}
    I_\nu\,[\mathrm{MJy\,sr}^{-1}] = \frac{F_\nu\,[\mu\mathrm{Jy}]}{23.504 \times (\theta/\mathrm{arcsec})^2}
\end{equation}
where $\theta$ is the pixel scale in arcseconds per pixel and the factor 23.504 follows from the solid-angle conversion between square arcseconds and steradians (1\,\MJysr\ $=$ 23.504\,$\mu$Jy\,arcsec$^{-2}$).

The wavelength offset between \wise W1 (3.4\,\micron) and \spitzer Ch1 (3.6\,\micron) is small. The Ch1 filter response spans $\Delta\lambda = 0.75$\,\micron\ centered at 3.58\,\micron\ \citep{Fazio2004}, and the broader W1 response overlaps it over most of that range \citep{Wright2010}. The two bandpasses therefore sample nearly the same rest-frame emission at any redshift; at $z \approx 0$, that emission is primarily the stellar continuum and the 3.3\,\micron\ PAH feature, with color terms of $<0.1$\,mag for most galaxy spectral energy distributions \citep[e.g.,][]{Jarrett2011}. We therefore neglect color corrections in this work.

\subsection{The COSMOS Field}
\label{sec:cosmos}

The Cosmic Evolution Survey \citep[\cosmos{};][]{Scoville2007} is a panchromatic imaging and spectroscopic survey of a 2\,deg$^2$ equatorial field centered at $(\alpha, \delta) = (150.1^\circ, +2.2^\circ)$, designed to probe galaxy evolution, large-scale structure, and dark matter through deep observations from X-ray to radio wavelengths.

The \cosmos field is well suited to training our super-resolution network for three reasons. The S-COSMOS program \citep{Sanders2007} and follow-up observations have produced deep \spitzer \irac imaging reaching $\approx 25.5$\,mag (AB, 3$\sigma$) at 3.6\,\micron. The AllWISE and \unwise programs \citep[e.g.,][]{Lang2014, Meisner2017} provide complete \wise coverage of the field. Extensive ancillary data, including photometric and spectroscopic redshifts, enable sample selection.

\begin{figure*}[!t]
\centering
\includegraphics[width=0.85\textwidth]{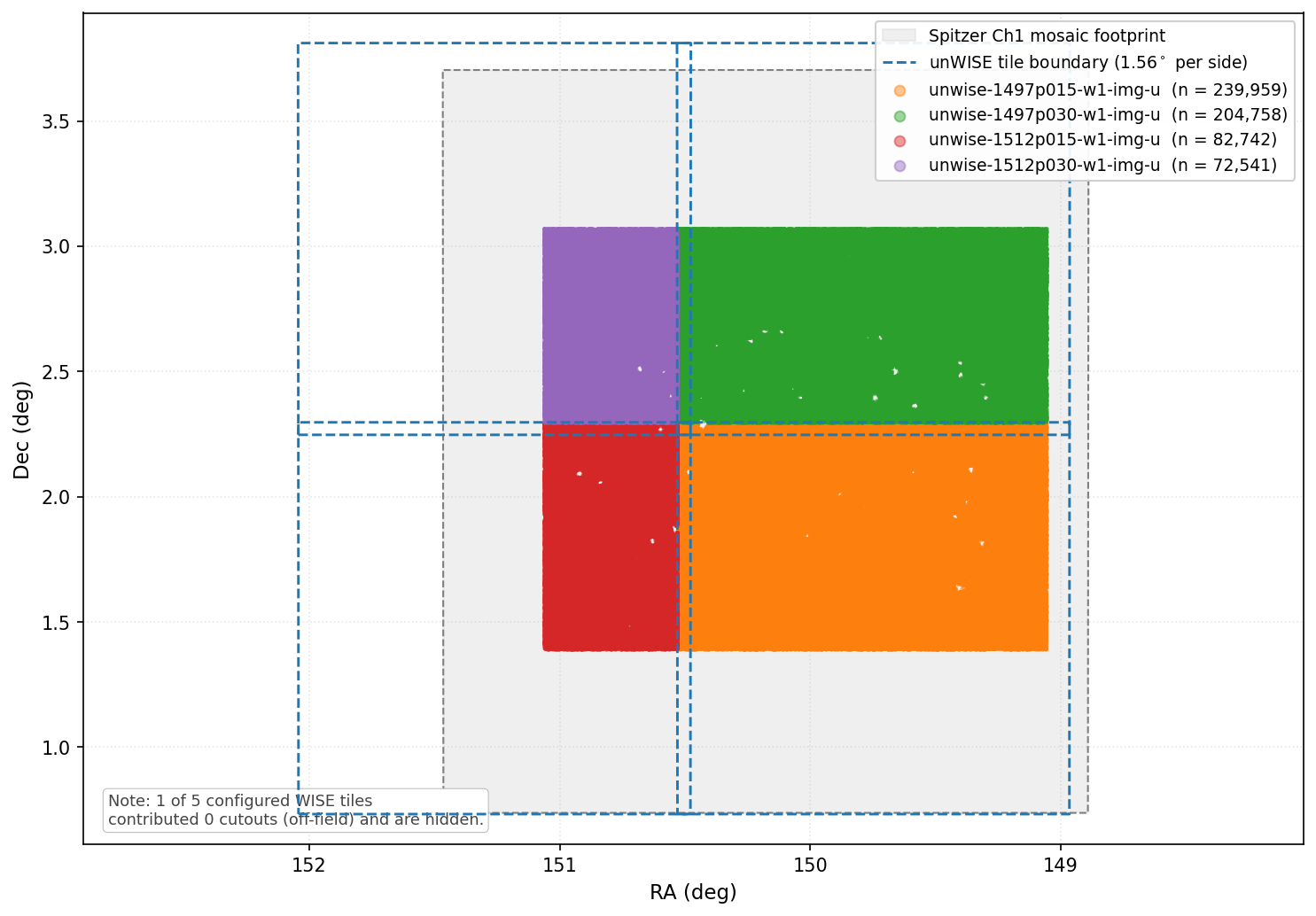}
\caption{Coverage of the \unwise W1 tiles contributing to the paired dataset, with per-tile counts of sampled catalog sources in the legend. The shaded region with the dashed gray outline marks the \spitzer-\cosmos Ch1 mosaic footprint; blue dashed boxes mark the 1.56$^\circ \times$1.56$^\circ$ boundaries of the four contributing \unwise tiles; points show the sampled source positions, colored by the tile that supplied the \wise cutout. A fifth configured tile lies outside the mosaic and contributed no cutouts. All training and test cutouts are drawn from the intersection of the tile coverage and the Spitzer footprint.}
\label{fig:tile_coverage}
\end{figure*}

We adopt the \spitzer-\cosmos Data Fusion mosaics \citep{Moneti2022, CosmicDawn2021}, which provide optimized 0.6\arcsec\ pixel$^{-1}$ \irac Ch1 imaging by combining all available \spitzer observations of the field with astrometric alignment and background matching. Figure~\ref{fig:tile_coverage} shows the spatial relationship between the five \unwise W1 tiles and the \spitzer-\cosmos Ch1 footprint.

We sample sources uniformly from the COSMOS2020\_CLASSIC photometric catalog \citep{Weaver2022, COSMOS2020classic2022} containing 1{,}720{,}700 unique sources (one catalog row per source) to construct our training sample. The only constraint applied is geometric: the source position must fall within the intersection of the \unwise W1 tile coverage and the \spitzer-\cosmos Ch1 mosaic, such that paired cutouts can be extracted. Each cutout is centered on a catalog source position. We do not apply photometric, classification, or magnitude cuts; the model trains on a diversity of sources (galaxies, stars, AGN) and brightnesses present in the catalog within the \wise/\spitzer overlap. Cutouts are therefore inherently mixed-content; each $38.4\arcsec \times 38.4\arcsec$ field of view contains many catalog sources of varied class and brightness, and we design the loss function to handle multisource cutouts independent of position (Section~\ref{sec:loss}).

\begin{figure*}[!t]
\centering
\includegraphics[width=0.95\textwidth]{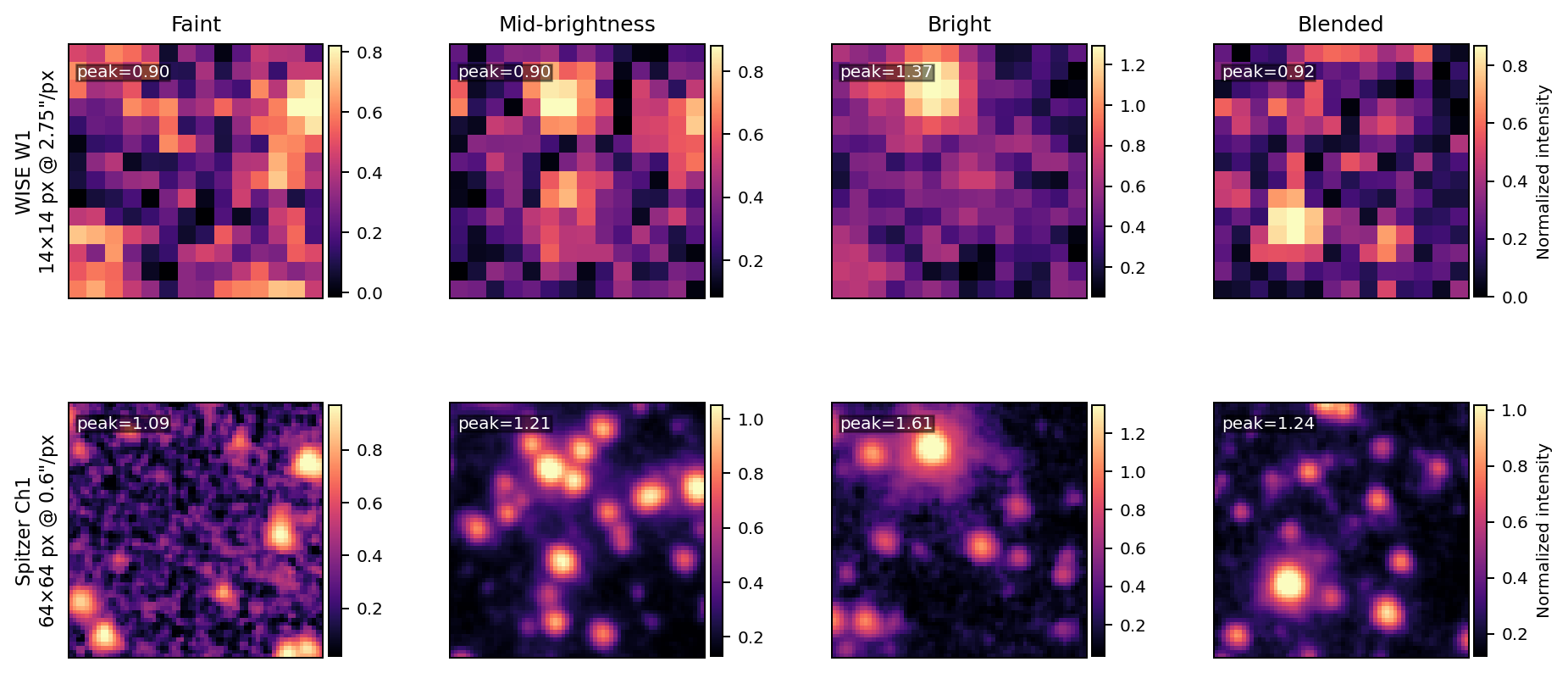}
\caption{Representative paired \wise$\to$~\spitzer cutouts from the test set, spanning faint, mid-brightness, bright, and blended regimes (peak intensities annotated; values in normalized units). Top: \wise W1 inputs (14$\times$14 pixels at 2.75\arcsec/pixel). Bottom: \spitzer \irac Ch1 ground truth (64$\times$64 pixels at 0.6\arcsec/pixel). Each pair covers the same 38.4\arcsec\ patch of sky. Each cutout is centered on a COSMOS2020 catalog source; because no brightness cut is applied (Section~\ref{sec:cosmos}), the central source need not be bright, nor the brightest object in the field. The \wise inputs show only 4--5 resolution elements across the cutout while the \spitzer truth resolves multiple distinct sources, illustrating the resolution gap the model must bridge. The crowded appearance is typical of the field (Section~\ref{sec:cosmos}).}
\label{fig:paired_examples}
\end{figure*}

For each sampled source, we extract paired cutouts centered on the source position: a $14\times14$ pixel \wise cutout ($\approx 38.5\arcsec \times 38.5\arcsec$) and a $64\times64$ pixel \spitzer cutout ($\approx 38.4\arcsec \times 38.4\arcsec$), covering the same physical area on the sky. The super-resolution factor is set by the ratio of the pixel scales ($2.75\arcsec/0.6\arcsec \approx 4.6$). Matching the sky coverage of the two arrays ensures the network learns to enhance resolution rather than extrapolate missing sky coverage. Figure~\ref{fig:paired_examples} shows representative paired cutouts spanning the brightness regimes encountered in the test set.

After removing samples containing invalid pixels (undefined or infinite values, primarily at mosaic edges), we obtain a uniformly sampled dataset of $\sim 557{,}000$ paired cutouts spanning the \cosmos catalog footprint. We split the dataset 70/15/15 (stratified by Spitzer truth peak brightness) into a training pool of 390{,}091, a validation set of 83{,}592 used to select the best model during training, and a held-out test set of 83{,}592 used exclusively for the metrics reported in Section~\ref{sec:results}. The split is stratified by the peak brightness of each cutout's Spitzer ground truth, defined as the brightest pixel of the $64\times64$ Spitzer cutout in normalized units. Cutouts are assigned to quartile bins at the 25th, 50th, and 75th percentiles of this peak brightness distribution, and the 70/15/15 split preserves the proportion of each quartile in every subset. The split fractions themselves carry no brightness meaning; stratification only guarantees that the training, validation, and test sets sample the same brightness distribution. The final reported model is trained on the full training pool (390{,}091 samples).

\begin{figure*}[!t]
\centering
\includegraphics[width=\textwidth]{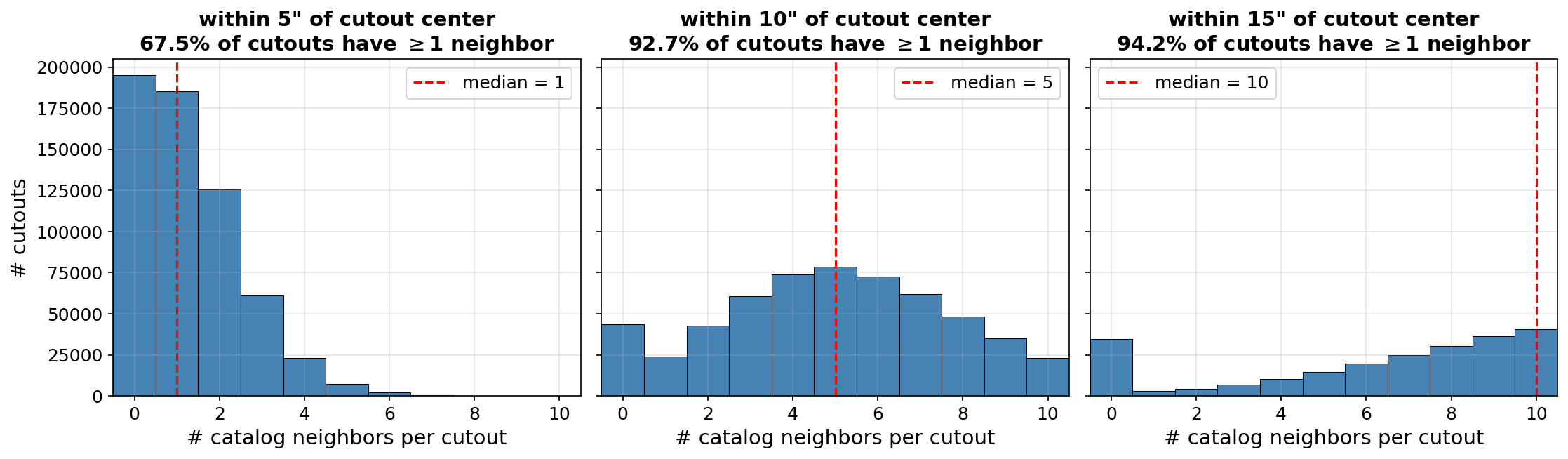}
\caption{COSMOS2020 catalog source density across the COSMOS field. Each panel is a histogram of the number of catalog neighbors per cutout, counted within 5\arcsec, 10\arcsec, and 15\arcsec\ of the cutout center; the panel titles give the fraction of cutouts with at least one neighbor and the dashed line marks the median count at that radius. The density is sufficient that the majority of 38.5\arcsec\ cutouts contain multiple catalog neighbors, motivating the position-agnostic loss design in Section~\ref{sec:loss}. Source counts use the quality-filtered catalog subset (IRAC Ch1 SNR\,$>$\,5, galaxies, MAG\,$<$\,25), the same selection as the companion analysis in Section~\ref{sec:companion}; the training sample itself applies no such cuts (Section~\ref{sec:cosmos}).}
\label{fig:catalog_density}
\end{figure*}

Deep \irac imaging of the \cosmos field yields a dense source catalog, and our cutouts are correspondingly crowded. Therefore, we train and evaluate the model in a regime where almost every cutout contains multiple bright sources within the $38.5\arcsec$ field of view. Figure~\ref{fig:catalog_density} shows the COSMOS2020 catalog source density across the field. This high source density motivates the deblending experiment in Section~\ref{sec:deblending} and the faint companion experiment in Section~\ref{sec:companion}.

\section{METHODS}
\label{sec:methods}

\subsection{Data Preprocessing}
\label{sec:preprocessing}

\begin{figure*}[!t]
\centering
\includegraphics[width=\textwidth]{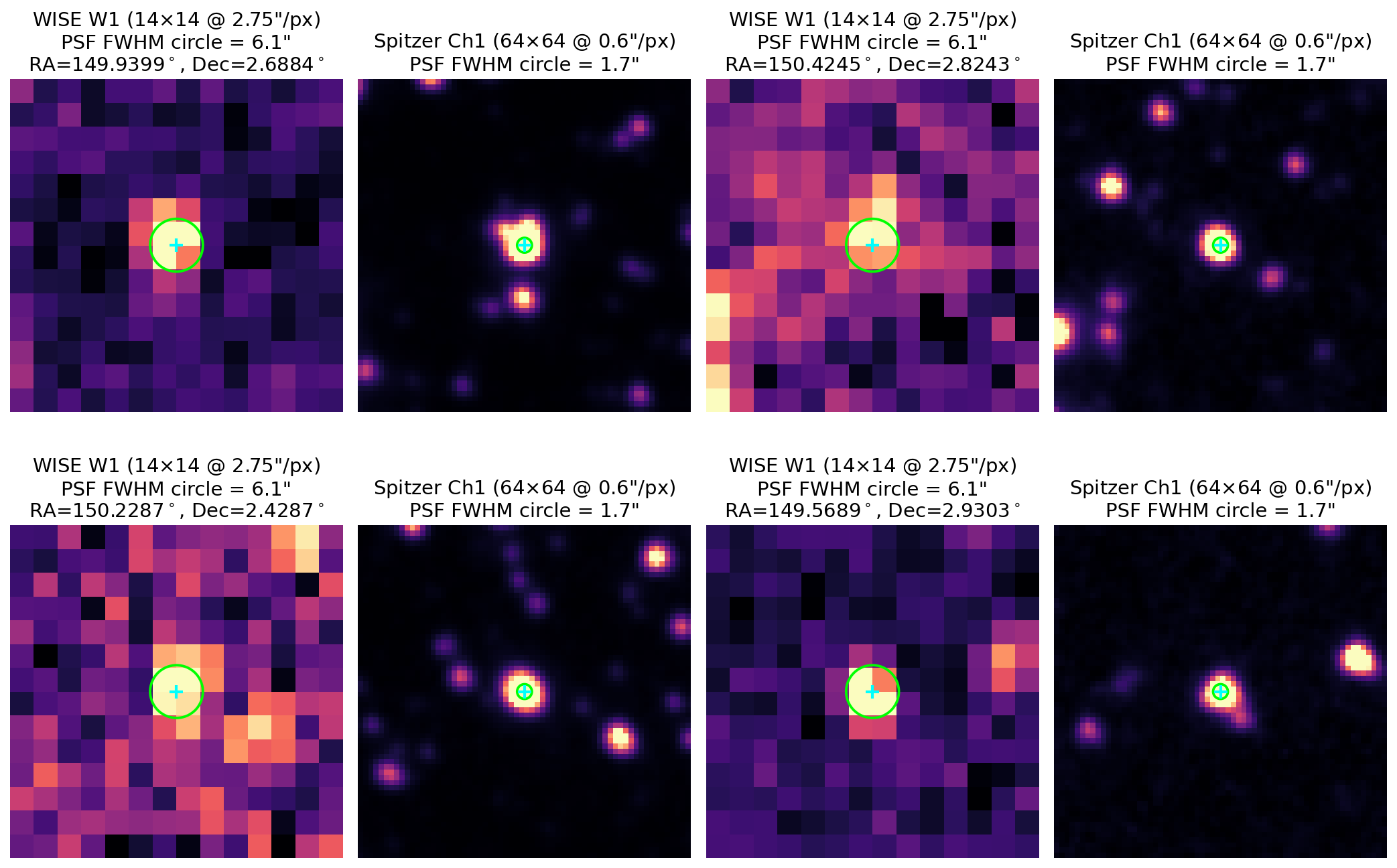}
\caption{Resolution gap between \wise W1 (3.4\,\micron, FWHM $\approx 6\arcsec$) and \spitzer \irac Ch1 (3.6\,\micron, FWHM $\approx 1.7\arcsec$) illustrated on four representative paired cutouts. The solid green circles mark the FWHM of each instrument's point-spread function, centered on the catalog position (cyan cross). The network must bridge this 3--4$\times$ difference in FWHM.}
\label{fig:psf_overlay}
\end{figure*}

Training a super-resolution network in this regime requires careful preprocessing. Our pipeline performs three steps: (i) conversion of the \wise and \spitzer data products to a common physical unit, (ii) compression of the large dynamic range typical of astronomical imaging, and (iii) preservation of physically meaningful negative pixel values. Figure~\ref{fig:psf_overlay} shows the resolution gap that the network must bridge.

\subsubsection{Unit Conversion}

The \unwise coadds are calibrated in linear Vega nanomaggies (nMgy) with a zero point of $\mathrm{MAGZP} = 22.5$ \citep{Lang2014, Meisner2017}, while \spitzer mosaics are in \MJysr surface-brightness units. Because the \unwise pixel values are already linear in flux, converting them to surface brightness is a single multiplicative rescaling:
\begin{equation}
    I_\nu\,[\mathrm{MJy\,sr}^{-1}] = C\,f\,[\mathrm{nMgy}],
\end{equation}
where the conversion constant $C = 3631 \times 10^{-0.4(\mathrm{MAGZP} + \Delta m_\mathrm{W1})} \times 10^{-6} / \Omega_\mathrm{pix} = 1.70 \times 10^{-3}$ combines the AB zero-point flux density of 3631\,Jy, the Vega-to-AB offset $\Delta m_\mathrm{W1} = 2.699$ for the W1 band \citep{Jarrett2011}, and the pixel solid angle $\Omega_\mathrm{pix} = (\theta_\mathrm{pix}/206265)^2$\,sr for a pixel scale of $\theta_\mathrm{pix} = 2.75$\arcsec.

The sky subtraction applied during the \unwise coaddition leaves a substantial fraction of pixels below zero ($\approx 43\%$ in our dataset; Section~\ref{sec:wise}). The rescaling above is linear and sign preserving, so these fluctuations carry through to \MJysr with their signs intact. We retain them rather than replacing them with a floor or a median value. Imputation would impose a deterministic positive offset on the background, place an artificial spike at the substituted value, and discard the evidence that a negative fluctuation carries against the presence of source flux at that position. After conversion the median positive \wise pixel is $6.1 \times 10^{-3}$\,\MJysr, against $2.4 \times 10^{-3}$\,\MJysr for the matched \spitzer cutouts, a ratio of 2.5.

\subsubsection{Asinh Normalization}

Astronomical images span several orders of magnitude in pixel value, from faint background to bright source cores. Standard normalization approaches such as min-max or z-score scaling either compress bright sources or amplify background noise. We adopt instead the inverse hyperbolic sine (asinh) stretch, which has become standard in astronomical visualization \citep{Lupton1999}:
\begin{equation}
    x_\mathrm{asinh} = \mathrm{asinh}\left(\frac{x}{x_\mathrm{soft}}\right)
\end{equation}
where $x_\mathrm{soft}$ is a softening parameter set to the median of positive pixel values. The asinh function behaves linearly for $|x| \ll x_\mathrm{soft}$ and logarithmically for $|x| \gg x_\mathrm{soft}$, compressing the dynamic range while preserving relative flux information. We note that asinh, unlike the logarithm, is defined for negative arguments and therefore handles background fluctuations natively. After the asinh transformation, we rescale values to approximately $[0, 1]$:
\begin{equation}
    x_\mathrm{norm} = \frac{x_\mathrm{asinh} - P_1}{P_{99} - P_1}
\end{equation}
where $P_1$ and $P_{99}$ are the 1st and 99th percentiles of the asinh-transformed pixel distribution. The percentiles are computed once over the full paired dataset rather than per cutout, so the same fixed transformation is applied to every image and relative brightness information between cutouts is preserved. The normalization parameters are stored and applied in inverse during evaluation, allowing recovery of physical flux units.

In brief, the network takes a $14\times14$ \wise cutout and produces a $64\times64$ image on the \spitzer pixel grid. It learns to do this from the $\sim 390{,}000$ matched \wise/\spitzer pairs: given many examples of how Spitzer resolution structure appears after blurring to \wise resolution, the network learns to invert that blurring. Three design choices matter most. The network examines the input at several spatial scales at once; during training it is penalized more heavily for errors on source pixels than on empty background; and it builds up the output resolution in stages rather than in a single jump. The remainder of this section summarizes these components. The full technical specification, including the architecture internals, the loss equations, and the training configuration, is given in Appendix~\ref{appendix:methods}.

\subsection{Neural Network Architecture}
\label{sec:architecture}

We develop an Enhanced Residual Channel Attention Network (Enhanced RCAN) optimized for astronomical super-resolution. The network is a deep stack of convolutional blocks with two features that matter here: skip connections at several levels, which let the training signal reach the early layers of a deep stack, and channel attention, which lets the network weight some feature maps more heavily than others rather than treating all of them equally (Appendix~\ref{appendix:methods}). Our architecture is built on the Residual Channel Attention Network of \citet{Zhang2018RCAN}, from which we inherit both, with the channel-attention module following \citet{Hu2018SENet}. We adopt subpixel upsampling \citep{Shi2016} at the output, as in the original RCAN.

Relative to the original RCAN, we make three architectural modifications and one hyperparameter retuning:
\begin{enumerate}
    \item \textbf{Multiscale input block} (Appendix~\ref{sec:multiscale}): three parallel input branches at different effective scales, fused via $1\times 1$ convolution, replacing RCAN's single $3\times 3$ input convolution. With only $\sim 4$ resolution elements across the WISE PSF in the $14\times 14$ input, the network benefits from receiving features at multiple spatial frequencies from the outset.
    \item \textbf{Multioutput progressive supervision} (Appendix~\ref{sec:upsampling}): three output heads at $28\times 28$, $56\times 56$, and $64\times 64$ with auxiliary loss weights $\lambda = 0.2, 0.3, 0.5$, replacing RCAN's single output at the target resolution. The intermediate supervision stabilizes training across the $4.6\times$ upsampling factor.
    \item \textbf{Source-focused composite loss} (Section~\ref{sec:loss}): weighted Huber + SSIM + spatial-gradient $L_1$ with position-agnostic source weighting, replacing RCAN's plain $L_1$ loss. Astronomical cutouts are dominated by background pixels; an unweighted loss would regress toward the local mean and discard the bright pixel structure that carries the science.
    \item \textbf{Hyperparameter retuning}: 4 residual groups $\times$ 8 RCABs ($=32$ blocks) at 128 base filters and 12.5M parameters, vs the RCAN default of 10 groups $\times$ 20 blocks at 64 base filters and $\sim 16$M parameters. Astronomical cutouts have far less semantic complexity than the natural-image DIV2K data RCAN was designed for; we trade depth for width, which captures the few-channel astronomical signal more efficiently.
\end{enumerate}

\begin{figure*}[!t]
\centering
\includegraphics[width=\textwidth]{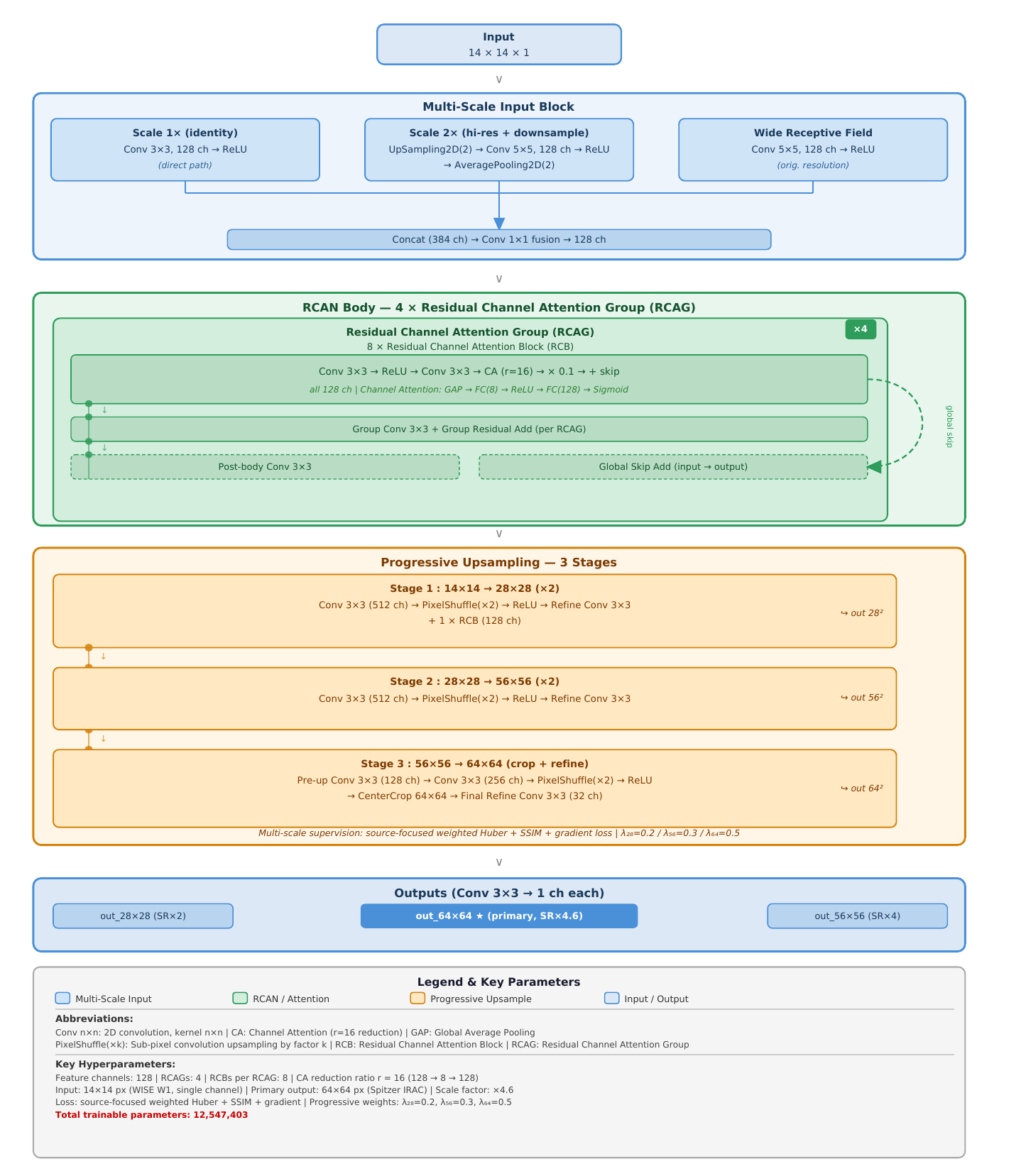}
\caption{
Architecture of the Enhanced Residual Channel Attention Network (Enhanced RCAN) used for super-resolution from \wise W1 ($14 \times 14$) to \spitzer IRAC Ch1 ($64 \times 64$). The model consists of a multiscale feature extraction block, four residual groups with channel attention (32 RCABs total), and progressive subpixel upsampling with intermediate supervision at 28$\times$28 and 56$\times$56 tap-offs.
}
\label{fig:architecture}
\end{figure*}

The network transforms a $14 \times 14 \times 1$ input to a $64 \times 64 \times 1$ output through the three stages shown in Figure~\ref{fig:architecture}. A multiscale feature extraction block applies parallel input convolutions at three effective scales, capturing fine detail and contextual information simultaneously. Stacked residual groups with channel attention then provide the learned feature representation. Finally, progressive upsampling raises the resolution gradually, with intermediate supervision at $28 \times 28$ and $56 \times 56$. The final model uses 128 base filters and 4 residual groups of 8 blocks each, for 12.5 million trainable parameters.

\subsection{Loss Function}
\label{sec:loss}

Standard loss functions such as mean squared error weight all pixels equally. This choice is suboptimal for astronomical images, where the scientifically important source pixels constitute a small fraction of the total and the loss is otherwise dominated by background. We therefore adopt a source-focused composite loss that emphasizes accurate reconstruction of source structure \citep{Hausen2020, Lauritsen2021}.

The loss combines three components. A weighted Huber term measures pixel fidelity. Pixels above a source threshold $\tau = 0.5$ in normalized units (approximately the median source flux) receive weight $w_\mathrm{source} = 3.0$, and background pixels receive $w_\mathrm{background} = 0.5$. An SSIM term encourages structurally accurate reconstruction, and a spatial-gradient term preserves sharp edges around sources. The components are combined with weights $0.5$, $0.35$, and $0.15$, applied at all three output scales with progressive supervision weights $0.2$, $0.3$, and $0.5$. The equations and the weight-setting procedure are given in Appendix~\ref{appendix:methods}.

\begin{figure*}[!t]
\centering
\includegraphics[width=0.95\textwidth]{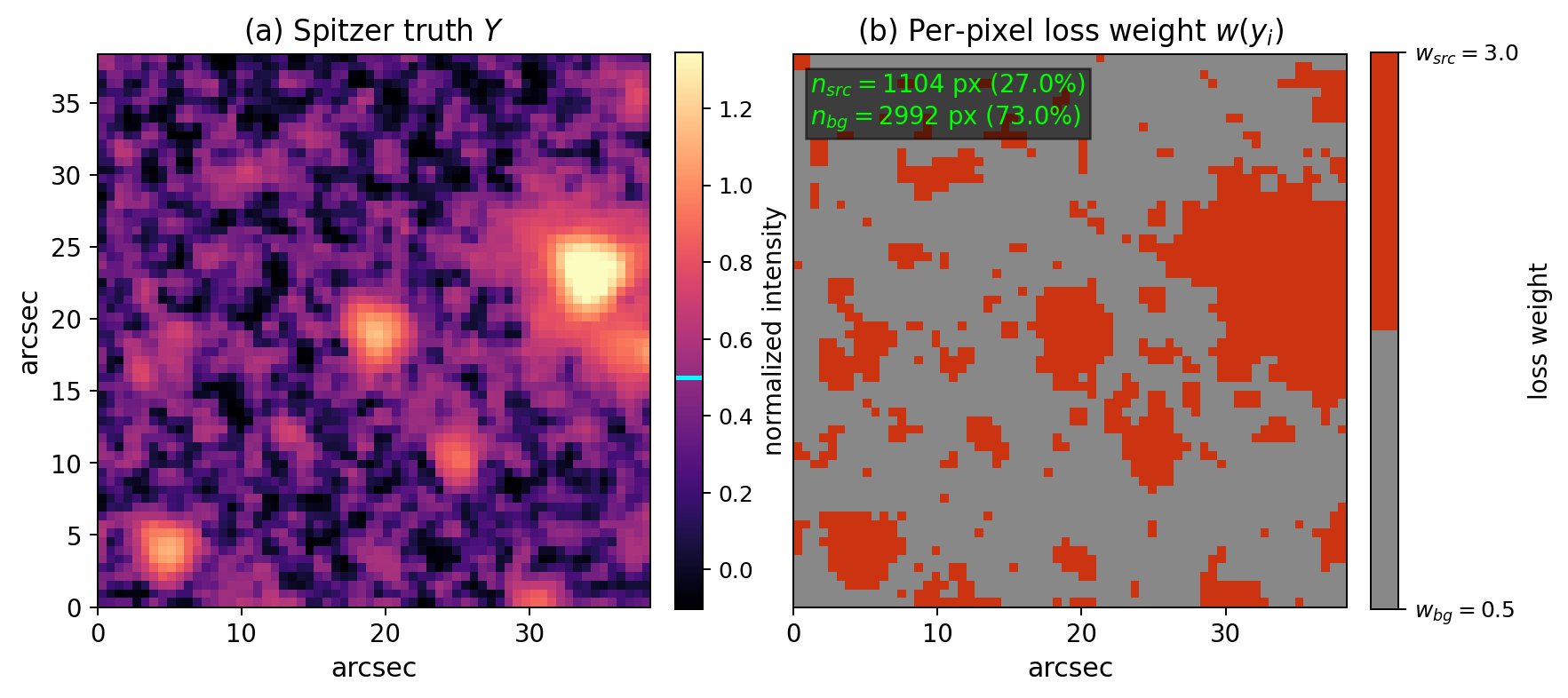}
\caption{The loss upweights pixels that belong to genuine sources relative to the background. (a) Spitzer truth for a representative test cutout in normalized intensity; the cyan line on the color bar marks the source threshold $\tau$. (b) Per-pixel loss weight $w(y_i)$: pixels above $\tau$ receive $w_\mathrm{source} = 3.0$ (red), all others receive $w_\mathrm{background} = 0.5$ (gray). The upweighted pixels cover $\approx 27\%$ of this cutout and are distributed across the field rather than concentrated at the cutout center, directly visualizing the position-agnostic property of the loss.}
\label{fig:loss_weight_map}
\end{figure*}

We emphasize that this weighting is position-agnostic: any pixel above the threshold $\tau$ receives $w_\mathrm{source}$ regardless of whether it lies at the cutout center or near the edge. Taken together with the cutout density characterized in Section~\ref{sec:cosmos} (the majority of cutouts contain multiple bright neighboring catalog sources), this means the model trains to reconstruct all bright pixels in the field rather than focusing on the central catalog primary. Figure~\ref{fig:loss_weight_map} illustrates the property on a representative test cutout. The position-agnostic weighting is relevant for interpreting the results binned by brightness in Section~\ref{sec:brightness} and the deblending experiment in Section~\ref{sec:deblending}, a point we return to below.

\subsection{Training Procedure}
\label{sec:training}

We train the network with the AdamW optimizer \citep{Loshchilov2019} and stop early when the validation loss no longer improves. The optimizer settings, learning-rate schedule, and hardware configuration are given in Appendix~\ref{appendix:methods}.

\paragraph{Training-set size.} We trained the network at 50{,}000, 100{,}000, and 200{,}000 samples to assess the sensitivity of model performance to dataset size. All headline metrics (SSIM, Pearson $r$, aperture peak and integrated flux errors, and the faint bin error) improved monotonically across the three sizes, and the gain from the 100k\,$\to$\,200k doubling exceeded the gain from the 50k\,$\to$\,100k doubling on six of nine metrics, indicating that the learning curve had not flattened at 200k. We therefore trained the final reported model on the full training pool ($\sim 390{,}000$ samples). The trend suggests that further gains are achievable with a larger training field.

\section{CHARACTERIZATION}
\label{sec:results}

\begin{figure*}[!t]
\centering
\includegraphics[width=\textwidth]{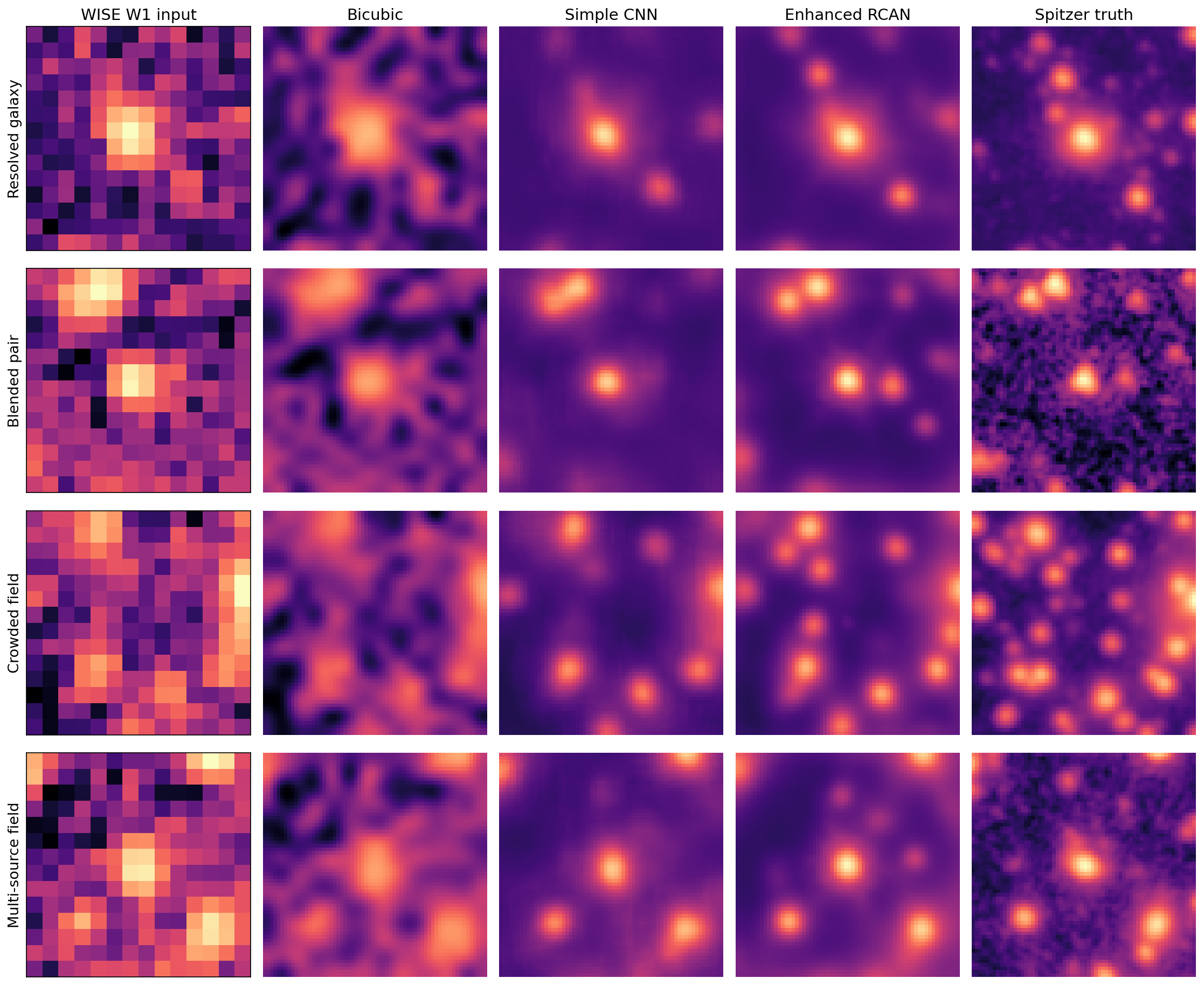}
\caption{Representative super-resolution results across the test set. Each row is one \wise$\to$~\spitzer test cutout, spanning a resolved galaxy, a blended close pair, a crowded field, and a moderate multi-source field. Columns show the \wise W1 input, the bicubic and Simple CNN baselines, the Enhanced RCAN output, and the \spitzer \irac Ch1 ground truth. The four $64\times64$ panels in each row share one display stretch. The Enhanced RCAN output recovers the compact sources of the Spitzer truth from the blended \wise input in every row. It separates the close pair in the second row and recovers the individual sources of the crowded field in the third. The baselines render the same fields as broad blends or recover only the brightest sources.}
\label{fig:sr_showcase}
\end{figure*}

We evaluate the trained model on the held-out test set of 83{,}592 paired cutouts and characterize its behavior along three axes: image quality, photometric fidelity, and source recovery. Figure~\ref{fig:sr_showcase} shows representative reconstructions across the test set alongside the baselines and the Spitzer ground truth. Throughout this section we compare the model against three baselines. Of these three baselines, bicubic and bilinear interpolation require no training. The Simple CNN baseline is a 4-layer SRCNN-style network trained from scratch on the same data with a mean squared error loss. We do not claim a single capability ready for blind deployment; rather, we report what the model does well and where it falls short. All metrics are computed in the asinh-normalized space defined in Section~\ref{sec:preprocessing} unless explicitly stated in physical units.

\subsection{Image Quality}
\label{sec:image_quality}

Table~\ref{tab:image_quality} summarizes whole-image quality metrics on the test set. We use structural similarity \citep[SSIM;][]{Wang2004} to compare two images through their local means, variances, and covariance, as defined below,
\begin{equation}
\mathrm{SSIM}(x,y) = \frac{(2\mu_x\mu_y + C_1)(2\sigma_{xy} + C_2)}{(\mu_x^2 + \mu_y^2 + C_1)(\sigma_x^2 + \sigma_y^2 + C_2)},
\end{equation}
where $\mu$, $\sigma^2$, and $\sigma_{xy}$ are the local means, variances, and covariance of the two images. $C_1$ and $C_2$ are stabilizing constants set by the dynamic range ($\mathrm{max\_val}$). The index is evaluated in $7\times7$-pixel windows and averaged over the image, and equals 1 for identical images. Image PSNR\,$= 20\log_{10}(\mathrm{max\_val}/\mathrm{RMSE})$ expresses pixel-level agreement on a logarithmic decibel scale, where higher is better. MAE and RMSE are the mean absolute and root-mean-square pixel differences, and Pearson $r$ is the linear correlation between predicted and true pixel values. SSIM and image PSNR are computed with $\mathrm{max\_val} = 2.0$, matched to the actual dynamic range of the asinh-normalized data (the bulk of pixels lie in $[0, 1]$ with a tail extending to $\sim 2.5$). We caution that larger choices of $\mathrm{max\_val}$ inflate these metrics artificially.

\begin{deluxetable}{lc}
\tablecaption{Image-Quality Metrics on the Test Set\label{tab:image_quality}}
\tablewidth{0pt}
\tablehead{
    \colhead{Metric} & \colhead{Value}
}
\startdata
SSIM\tablenotemark{a} & $0.51 \pm 0.14$ \\
Image PSNR\tablenotemark{b} & $23.6 \pm 1.6$ dB \\
MAE\tablenotemark{c} & $0.100$ \\
RMSE\tablenotemark{c} & $0.137$ \\
Pearson $r$\tablenotemark{d} & $0.75$ \\
\enddata
\tablenotetext{a}{Structural Similarity Index at $\mathrm{max\_val} = 2.0$; mean $\pm$ 1$\sigma$ over 83,592 per-cutout values.}
\tablenotetext{b}{Peak Signal-to-Noise Ratio at $\mathrm{max\_val} = 2.0$; same averaging as note $a$.}
\tablenotetext{c}{Aggregated over all pixels in 83,592 test cutouts in normalized space.}
\tablenotetext{d}{Per-pixel correlation between predicted and Spitzer truth output, aggregated over the test set.}
\end{deluxetable}

Structural fidelity is moderate (SSIM\,$=$\,0.51), sufficient for visual identification of sources and large-scale morphology but limited at the fine-structure scale. The per-pixel correlation ($r = 0.75$) is reasonable but inflated by the large fraction of background pixels. We note that these image quality metrics measure how well the model reproduces the asinh-stretched intensity field of Section~\ref{sec:preprocessing}, a stretch adopted both to compress the large dynamic range of astronomical images and because it is defined for the negative sky-subtracted pixels that a logarithm cannot represent. They do not measure photometric fidelity directly, which we report next.

\subsection{Photometric Accuracy}
\label{sec:photometry}

We report the metrics related to photometric accuracy under two complementary definitions so that the qualitative finding is robust to the choice of metric. The first is a threshold mask definition, commonly used in pixel-wise super-resolution evaluation. The second is aperture photometry at the central catalog source position (Table~\ref{tab:photometry_aperture}), which is the standard astronomical photometry methodology.

Under the threshold mask definition, the peak flux error compares the brightest pixel of the predicted and ground truth cutouts. The relative error is computed per cutout in normalized space and has a mean of $7.6\%$ and a median of $6.6\%$ over the 83,592 test cutouts. The integrated flux error is computed the same way for the flux summed over a source mask. The mask is the set of pixels whose ground truth normalized value exceeds the source threshold $\tau = 0.5$. It is derived from the truth image alone and applied identically to prediction and truth. No catalog-based segmentation is involved. This error has a mean of $20.4\%$ and a median of $19.1\%$ on the same test set. It is biased upward by the model's oversmoothing of source profiles (Section~\ref{sec:failure_mode}). The predictions also preserve the signal-to-noise ratio of the truth images, with a predicted-to-truth SNR ratio of $0.97$. The SNR of each image is the mean over the source mask pixels divided by the standard deviation of the pixels outside the mask. Both flux metrics depend on how the source is defined. The brightest pixel may belong to a contaminating neighbor in our crowded cutouts, and the integrated flux sums pixels above an arbitrary threshold. We therefore complement these threshold-based measurements with aperture photometry below.

\begin{deluxetable*}{lcc}
\tablecaption{Aperture Photometry at the Central Catalog Source\label{tab:photometry_aperture}}
\tablewidth{0pt}
\tablehead{
    \colhead{Method} & \colhead{Peak aperture\tablenotemark{a}} & \colhead{Integrated aperture\tablenotemark{b}}
}
\startdata
Bicubic & $32.0\%$ & $22.5\%$ \\
Bilinear & $30.2\%$ & $21.7\%$ \\
Simple CNN (4 layers) & $27.1\%$ & $16.8\%$ \\
Enhanced RCAN (this work) & 20.1\% & 10.8\% \\
\enddata
\tablenotetext{a}{Median relative error in aperture flux at $r = 2$ px ($\approx 1.2\arcsec$, $\approx 0.7\times$ Spitzer FWHM); aperture centered at the cutout center, where the COSMOS catalog primary sits by construction.}
\tablenotetext{b}{Median relative error in aperture flux at $r = 6$ px ($\approx 3.6\arcsec$, $\approx 2\times$ Spitzer FWHM); same centering as note $a$. No background subtraction (see caveat in text).}
\end{deluxetable*}

\begin{figure*}[!t]
\centering
\includegraphics[width=\textwidth]{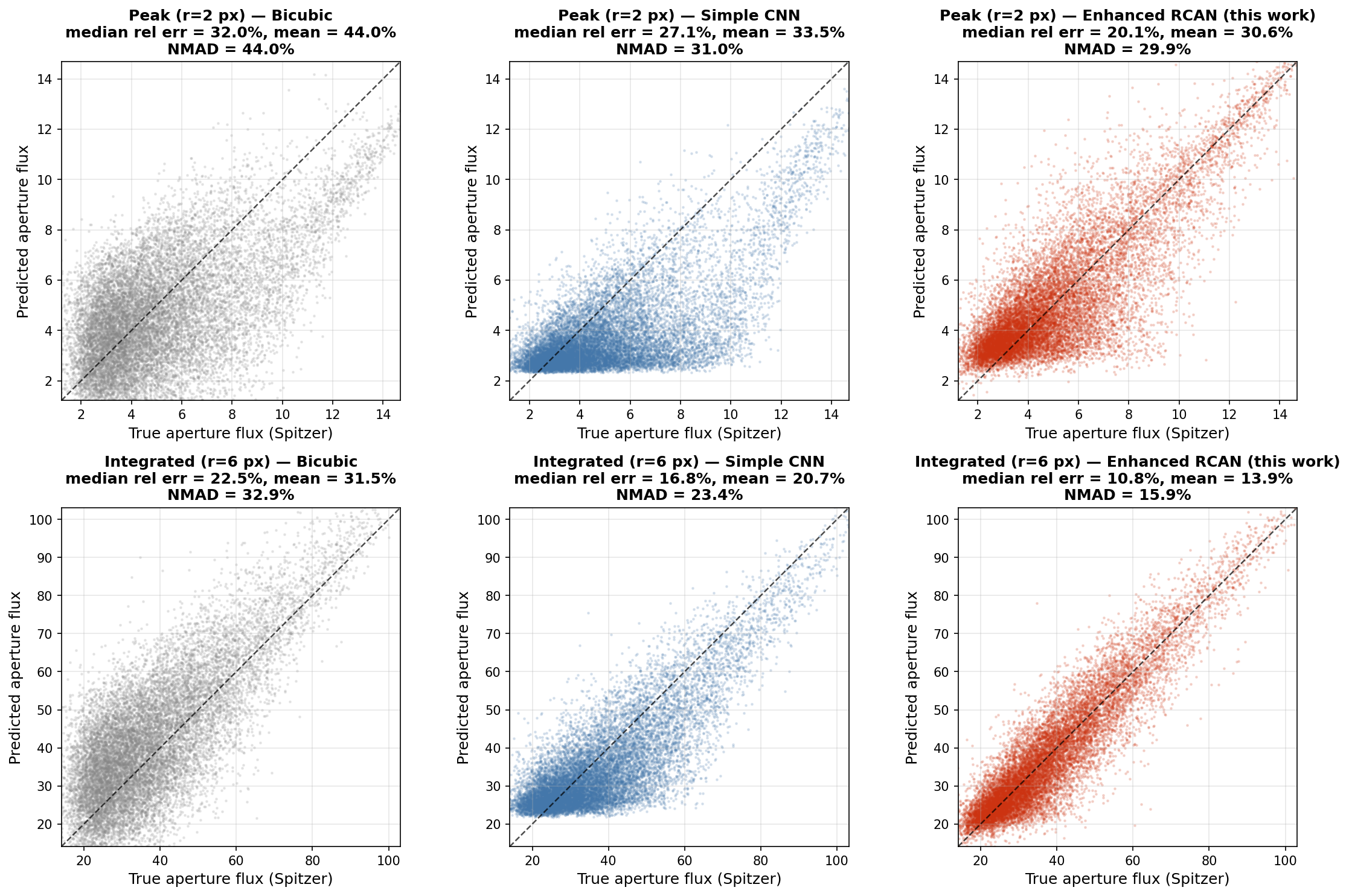}
\caption{Aperture photometry at the central catalog position: predicted (y-axis) versus Spitzer ground truth (x-axis) for each method. Top: peak aperture ($r = 2$ px). Bottom: integrated aperture ($r = 6$ px). The dashed line in each panel marks the 1:1 relation. The Enhanced RCAN (right column) follows the 1:1 line tightly. The vertical scatter differs substantially between methods. We quantify the dispersion with the normalized median absolute deviation of the signed relative error, defined as 1.4826 times the median absolute deviation from the median and annotated in each panel as NMAD. The Enhanced RCAN shows both the smallest dispersion, with an integrated aperture NMAD of $15.9\%$ versus $32.9\%$ for bicubic, and the smallest bias, with a signed median error of $+0.7\%$ versus $+11.0\%$ for bicubic.}
\label{fig:aperture_scatter}
\end{figure*}

Aperture photometry confirms the threshold mask finding. The Enhanced RCAN recovers the central source's aperture peak flux to a median relative error of $20\%$ and the integrated aperture flux to $11\%$, both a factor of $\approx 1.5$--$2$ better than the interpolation baselines (Section~\ref{sec:brightness}). The integrated aperture error of RCAN ($10.8\%$) is in fact lower than its threshold mask integrated error of $\approx 19\%$, the median relative error of the flux summed over all pixels above the source threshold $\tau$. The fixed radius aperture isolates the central source. The threshold mask is contaminated by neighbors in our crowded cutouts (Section~\ref{sec:cosmos}). The Simple CNN baseline exhibits a subtler pattern: its aperture peak error ($27\%$) is modestly better than bicubic's ($32\%$), and its aperture integrated error ($16.8\%$) is second only to RCAN. The Simple CNN's threshold mask integrated error ($30.4\%$), by contrast, is the worst of any method, because the oversmoothed PSF wings inflate the flux summed over the source mask even when the brightest pixel is correctly placed. Figure~\ref{fig:aperture_scatter} demonstrates these results.

\paragraph{Caveat on background subtraction.} Standard astronomical aperture photometry subtracts a sky background estimated from an annulus surrounding the source. We tested this convention with an annulus at radii [8, 12] px and found that for all methods (including Spitzer ground truth as a sanity check), the background-subtracted aperture error rose above $30\%$. The cause is the source density of the COSMOS field (Section~\ref{sec:cosmos}): the background annulus is contaminated by neighboring sources, such that subtraction of its median overcorrects. In crowded fields the background is therefore usually estimated in other ways, from a global background map or by fitting overlapping sources simultaneously \citep[e.g.,][]{Bertin1996}. Because our purpose here is a differential comparison rather than absolute photometry, we report non-subtracted aperture measurements: the same aperture is applied to each method's output and to the Spitzer truth, so the residual background enters both terms alike. Users who apply the model to dense fields should still account for the background using one of these methods.

\subsection{Brightness Dependence}
\label{sec:brightness}

\begin{figure*}[!t]
\centering
\includegraphics[width=0.95\textwidth]{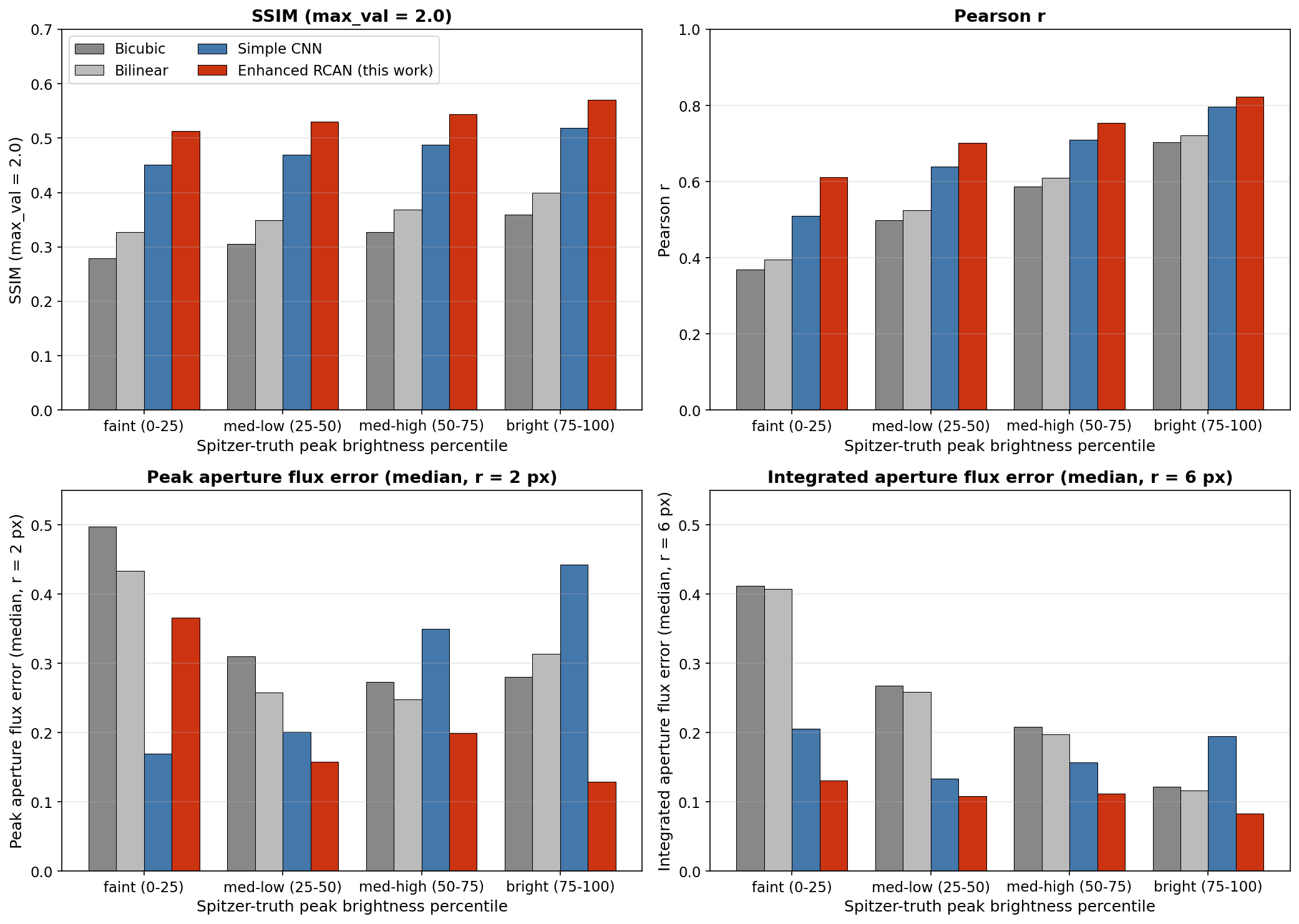}
\caption{Metrics binned by brightness across the four methods. Each panel shows one metric, with one bar group per brightness bin and one bar per method. The flux panels show the median relative error in aperture flux at the central catalog position, at $r = 2$ px for the peak and $r = 6$ px for the integrated aperture, the same definitions as Tables~\ref{tab:brightness_rcan} and \ref{tab:brightness_all}. The RCAN advantage is largest in the faint and intermediate bins, and RCAN keeps the lowest integrated flux error in every bin, including the brightest quartile at $8.3\%$ versus $12.2\%$ for bicubic.}
\label{fig:brightness_binned}
\end{figure*}

The population-averaged metrics in Sections~\ref{sec:image_quality}--\ref{sec:photometry} mask a strong, monotonic dependence of model performance on source brightness. We bin the test set into quartiles by Spitzer truth peak brightness and report per-bin metrics. Table~\ref{tab:brightness_rcan} shows the Enhanced RCAN performance, Table~\ref{tab:brightness_all} expands to all four methods, and Figure~\ref{fig:brightness_binned} visualizes the same data.

\begin{deluxetable*}{lccccc}
\tablecaption{RCAN Brightness Dependence\label{tab:brightness_rcan}}
\tablewidth{0pt}
\tablehead{
    \colhead{Brightness bin\tablenotemark{a}} & \colhead{SSIM\tablenotemark{b}} & \colhead{Pearson $r$\tablenotemark{c}} & \colhead{Image PSNR\tablenotemark{b}} & \colhead{Peak err\tablenotemark{d}} & \colhead{Int.\ flux err\tablenotemark{e}}
}
\startdata
Faint (0--25\%) & 0.48 & 0.61 & 23.7 dB & $36.6\%$ & $13.1\%$ \\
Med-low (25--50\%) & 0.50 & 0.70 & 23.5 dB & $15.7\%$ & $10.8\%$ \\
Med-high (50--75\%) & 0.52 & 0.75 & 23.5 dB & $19.9\%$ & $11.2\%$ \\
Bright (75--100\%) & 0.55 & 0.82 & 23.8 dB & 12.9\% & 8.3\% \\
\enddata
\tablenotetext{a}{Cutouts binned by Spitzer truth peak brightness percentile; $n \approx 20{,}898$ per bin.}
\tablenotetext{b}{Mean over per-cutout values at $\mathrm{max\_val} = 2.0$.}
\tablenotetext{c}{Per-pixel correlation between predicted and Spitzer truth output, averaged over per-cutout values.}
\tablenotetext{d}{Median relative error in aperture flux at $r = 2$ px ($\approx 0.7\times$ Spitzer FWHM).}
\tablenotetext{e}{Median relative error in aperture flux at $r = 6$ px ($\approx 2\times$ Spitzer FWHM).}
\end{deluxetable*}

\begin{deluxetable*}{llcccc}
\tablecaption{Brightness Dependence Across Methods\label{tab:brightness_all}}
\tablewidth{0pt}
\tablehead{
    \colhead{Bin\tablenotemark{a}} & \colhead{Method} & \colhead{SSIM\tablenotemark{b}} & \colhead{Pearson $r$\tablenotemark{c}} & \colhead{Peak err\tablenotemark{d}} & \colhead{Int.\ flux err\tablenotemark{e}}
}
\startdata
Faint   & Bicubic    & 0.24 & 0.37 & $49.8\%$ & $41.2\%$ \\
        & Bilinear   & 0.29 & 0.39 & $43.4\%$ & $40.7\%$ \\
        & Simple CNN & 0.41 & 0.51 & 17.0\% & $20.6\%$ \\
        & RCAN & 0.48 & 0.61 & $36.6\%$ & 13.1\% \\
\hline
Med-low & Bicubic    & 0.27 & 0.50 & $31.0\%$ & $26.7\%$ \\
        & Bilinear   & 0.31 & 0.52 & $25.8\%$ & $25.9\%$ \\
        & Simple CNN & 0.44 & 0.64 & $20.1\%$ & $13.3\%$ \\
        & RCAN & 0.50 & 0.70 & 15.7\% & 10.8\% \\
\hline
Med-high & Bicubic    & 0.30 & 0.59 & $27.3\%$ & $20.8\%$ \\
        & Bilinear   & 0.34 & 0.61 & $24.8\%$ & $19.8\%$ \\
        & Simple CNN & 0.46 & 0.71 & $34.9\%$ & $15.7\%$ \\
        & RCAN & 0.52 & 0.75 & 19.9\% & 11.2\% \\
\hline
Bright  & Bicubic    & 0.33 & 0.70 & $28.0\%$ & $12.2\%$ \\
        & Bilinear   & 0.37 & 0.72 & $31.4\%$ & $11.6\%$ \\
        & Simple CNN & 0.49 & 0.80 & $44.3\%$ & $19.5\%$ \\
        & RCAN & 0.55 & 0.82 & 12.9\% & 8.3\% \\
\enddata
\tablenotetext{a}{Cutouts binned by Spitzer truth peak brightness percentile; $n \approx 20{,}898$ per bin (same definition as Table~\ref{tab:brightness_rcan}).}
\tablenotetext{b}{Mean over per-cutout values at $\mathrm{max\_val} = 2.0$.}
\tablenotetext{c}{Per-pixel correlation between predicted and Spitzer truth output, averaged over per-cutout values.}
\tablenotetext{d}{Median relative error in aperture flux at $r = 2$ px.}
\tablenotetext{e}{Median relative error in aperture flux at $r = 6$ px.}
\end{deluxetable*}

RCAN performance has a strong monotonic dependence on source brightness. The aperture integrated flux error decreases from $13\%$ on the faintest 25\% of sources to $8\%$ on the brightest 25\%. The Pearson $r$ improves from 0.61 to 0.82, and SSIM improves from 0.48 to 0.55. For applications targeting bright sources only, the model performs substantially better than the population-averaged headline numbers in Section~\ref{sec:photometry} suggest.

The model's advantage over the baselines is largest on the brightest sources. On the brightest quartile, the RCAN aperture peak error ($12.9\%$) is a factor of $\approx 2$ better than bicubic ($28.0\%$) and $\approx 3$ better than Simple CNN ($44.3\%$); the RCAN integrated flux error ($8.3\%$) is $\approx 1.5\times$ better than bicubic. The faint quartile is more nuanced: RCAN dominates the integrated flux metric ($13\%$ vs $41\%$ for bicubic, a factor of $\approx 3$ improvement), but Simple CNN achieves a lower aperture peak error ($17\%$) than RCAN ($37\%$). The latter result is plausibly an artifact of training set diversity: when trained on the unfiltered catalog (which contains stars, AGN, and faint subdetection sources), Simple CNN's aggressive oversmoothing suppresses noise more effectively than RCAN's sharper reconstruction in the faint, noise-dominated regime, at the cost of destroying peak structure on bright sources. Two implications follow. First, the trained model adds the most value precisely where it matters most: bright sources, where photometry is the central scientific quantity. Second, faint source aperture peak measurements remain a hard problem. The choice of method depends on whether one prefers the smooth, low-noise output of the Simple CNN or the sharper but noisier output of RCAN. The Simple CNN aperture peak error in fact worsens with brightness ($17\%$ at faint, $44\%$ at bright); this is the signature of the oversmoothing produced by training objectives that minimize mean squared error \citep[e.g.,][]{Ledig2017, Zhao2017}, destroying source compactness in the regime where the source has the most flux to redistribute.

\subsection{Source Deblending}
\label{sec:deblending}

The training and test data are dominated by crowded multisource cutouts (Section~\ref{sec:cosmos}). The model has therefore implicitly trained on the \wise\,$\to$\,\spitzer mapping in a regime where multiple comparably bright sources sit inside each cutout, with the source-focused loss weighting bright pixels independent of position (Section~\ref{sec:loss}). A direct measurement of the deblending capability is therefore well-defined.

\paragraph{Experimental design.} For each of 83{,}592 test cutouts, we run a simple peak detector (local maxima above a threshold with a minimum spacing constraint) on each method's output (bicubic upsample, bilinear, Simple CNN, Enhanced RCAN) and on the Spitzer ground truth. We use a simple peak detector rather than a full source extraction pipeline \citep[e.g., SExtractor;][]{Bertin1996} so that the comparison measures the quality of each method's reconstructed images rather than the behavior of the extraction software. We adopt the Spitzer truth peaks as an oracle, in the sense that peaks detected in Spitzer truth define what is detectable at Spitzer resolution. For each method, we match each detected peak to the nearest Spitzer truth peak within a $1.5\arcsec$ tolerance ($\approx 1$ Spitzer FWHM), and compute the recall (fraction of Spitzer truth peaks recovered), precision (fraction of method peaks that match), median position error, and median flux error on matched peaks. The results are binned by minimum truth peak separation per cutout, with the 3--5\arcsec\ bin corresponding to the separations where \wise\ blends sources that \spitzer\ separates.

\begin{deluxetable*}{lccccc}
\tablecaption{Deblending Performance: Overall (a) and by Peak Separation (b)\label{tab:deblending}}
\tablewidth{0pt}
\tablehead{
    \colhead{(a) Overall\tablenotemark{a}} & \colhead{Recall\tablenotemark{b}} & \colhead{Precision\tablenotemark{c}} & \colhead{Cardinality match\tablenotemark{d}} & \colhead{Pos.\ err\tablenotemark{e}} & \colhead{Flux err\tablenotemark{f}}
}
\startdata
Bicubic & $10.3\%$ & $59.6\%$ & $12.2\%$ & 1.41 px & $16.2\%$ \\
Bilinear & $7.6\%$ & $52.2\%$ & $11.4\%$ & 1.41 px & $17.2\%$ \\
Simple CNN & $24.5\%$ & $75.4\%$ & $17.8\%$ & 1.50 px & $10.7\%$ \\
Enhanced RCAN (this work) & 39.5\% & 69.2\% & 30.2\% & 1.50 px & 5.8\% \\
\hline
\colhead{(b) Recall by separation\tablenotemark{g}} & \colhead{$n$ cutouts} & \colhead{Bicubic} & \colhead{Simple CNN} & \colhead{RCAN} & \colhead{Improvement\tablenotemark{h}} \\
\hline
2--3\arcsec & 559 & $8.3\%$ & $22.1\%$ & $35.5\%$ & 4.3$\times$ \\
3--5\arcsec\ (deblending regime) & 9{,}285 & 9.2\% & 21.4\% & 35.4\% & 3.8$\times$ \\
5--10\arcsec & 20,156 & $10.6\%$ & $23.8\%$ & $40.7\%$ & 3.8$\times$ \\
10--64\arcsec & 31,365 & $10.7\%$ & $26.3\%$ & $40.4\%$ & 3.8$\times$ \\
\enddata
\tablenotetext{a}{Per-method peak recovery statistics across 83,592 cutouts (229,857 Spitzer truth peaks total). Peak threshold 1.0 in normalized space ($\approx$ 99.9th percentile of $Y_{\mathrm{test}}$); min spacing 3 px ($\approx 1.8\arcsec$); match tolerance 2.5 px ($\approx 1.5\arcsec$).}
\tablenotetext{b}{Fraction of Spitzer truth peaks within 1.5\arcsec\ of a peak detected by the method.}
\tablenotetext{c}{Fraction of peaks detected by each method that match a Spitzer truth peak.}
\tablenotetext{d}{Fraction of cutouts where the method's predicted peak count equals Spitzer truth's.}
\tablenotetext{e}{Median pixel distance between matched method-peak and Spitzer truth peak.}
\tablenotetext{f}{Median relative flux error on matched peaks.}
\tablenotetext{g}{Cutouts binned by minimum pairwise separation of Spitzer truth peaks. The 3--5\arcsec\ bin is the regime where \wise\ blends sources that \spitzer\ separates.}
\tablenotetext{h}{Ratio of RCAN recall to bicubic recall in this separation bin.}
\end{deluxetable*}

\begin{figure*}[!t]
\centering
\includegraphics[width=0.8\textwidth]{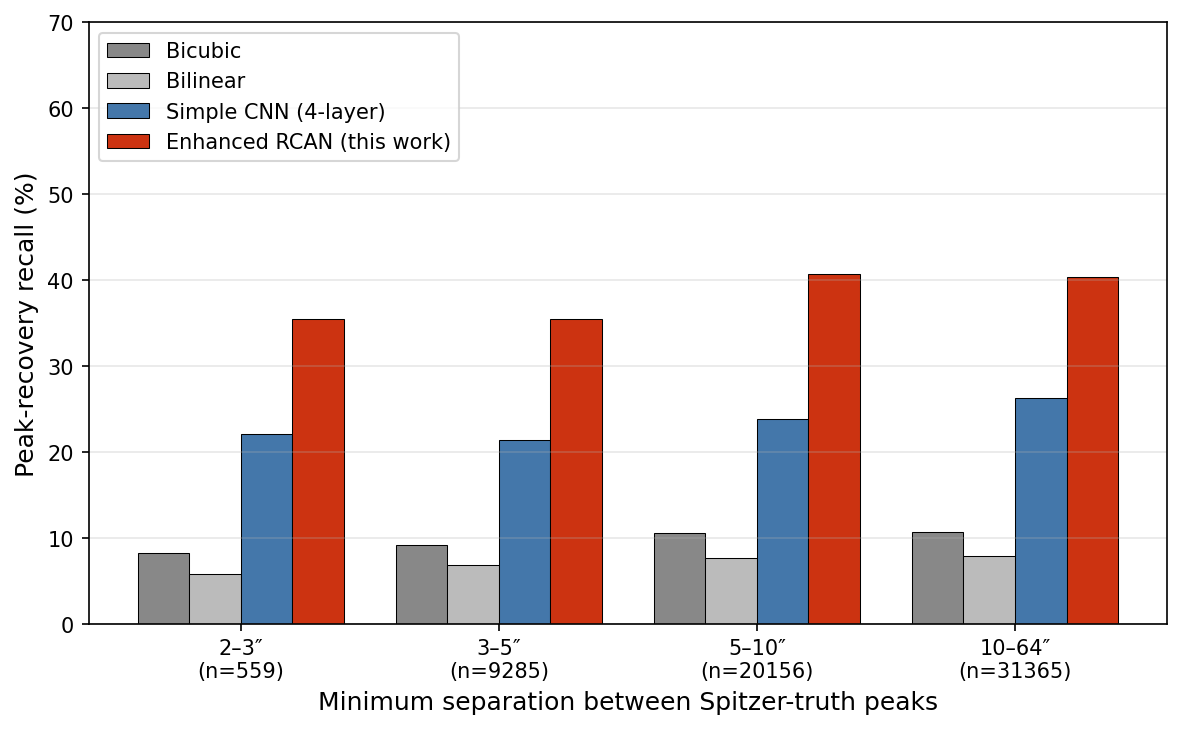}
\caption{Deblending peak recovery recall versus minimum source separation, for the four methods. The Enhanced RCAN consistently recovers $\approx 4\times$ more of the peaks detectable in the Spitzer truth than bicubic upsampling across the deblending regime (2--10\arcsec), indicating that the model performs real source separation rather than interpolating WISE PSF wings.}
\label{fig:deblending_curve}
\end{figure*}

At 3--5\arcsec\ separations, the trained model recovers $35.4\%$ of the source peaks detectable in the Spitzer truth compared with $9.2\%$ for bicubic upsampling, a factor of $\approx 3.8$ improvement (Table~\ref{tab:deblending} and Figure~\ref{fig:deblending_curve}). Across all separation bins, the model improves the recall by $\approx 4\times$ relative to bicubic, with excellent flux accuracy on matched peaks (median $5.8\%$ relative error). The implication is that the model is performing real deblending, not merely preserving artifacts of the \wise\ PSF wings through interpolation.

\paragraph{Caveats.} The model misses 60\% of the peaks detectable in the Spitzer truth across the full test set, with the misses distributed across truth peak counts and separations. The oversmoothing failure mode (Section~\ref{sec:failure_mode}) accounts for a substantial fraction of these misses regardless of separation. The precision is 69\%, such that $\approx 31\%$ of peaks predicted by the model do not match a Spitzer truth peak; some of these are spurious, and some are real Spitzer peaks just outside the $1.5\arcsec$ matching tolerance. We caution that the peak detection threshold (1.0 in normalized space) is a methodological choice, though the qualitative finding (model $\approx 4\times$ better than bicubic) is robust to threshold variation.

\subsection{Faint Companion Detection}
\label{sec:companion}

We test the ability of the model to recover known catalog companions in the COSMOS field. For each test cutout, the central catalog source is the primary (always at the cutout center, by construction); any other COSMOS2020 catalog source within $15\arcsec$ at flux ratio $<1.0$ is a companion. We project each companion's (RA, Dec) into the cutout pixel grid and check whether each method produces a peak (using a permissive threshold of 0.3 in normalized space, with a min spacing of 3 px) within $1.5\arcsec$ of the projected position. Across 83{,}592 test cutouts, there are 1{,}029{,}115 catalog companions in total.

\paragraph{Spitzer ceiling.} Spitzer ground truth itself produces a peak above the 0.3 threshold for only 63.5\% of catalog companions; the remaining $\sim 37\%$ are too faint or too close to the central source to register as distinct peaks in the actual Spitzer cutout. This fraction is the upper bound for any method's recall in this setup.

\begin{deluxetable*}{lcc}
\tablecaption{Faint Companion Recall Against Catalog Truth\label{tab:companion}}
\tablewidth{0pt}
\tablehead{
    \colhead{Method} & \colhead{Overall recall\tablenotemark{a}} & \colhead{Normalized recall\tablenotemark{b}}
}
\startdata
Bicubic & $12.8\%$ & $20.1\%$ \\
Bilinear & $12.0\%$ & $19.0\%$ \\
Simple CNN (4 layers) & $7.9\%$ & $12.4\%$ \\
Enhanced RCAN (this work) & 16.0\% & 25.2\% \\
\textit{Spitzer truth (oracle)} & \textit{63.5\%} & \textit{100.0\%} \\
\enddata
\tablenotetext{a}{Fraction of catalog companions whose projected pixel position lies within 1.5\arcsec\ of a peak in the method's output. Catalog companions: 1{,}029{,}115 across all 83{,}592 test cutouts.}
\tablenotetext{b}{Overall recall divided by the Spitzer truth ceiling (63.5\%); the truth value sets the upper bound on what is detectable in this setup.}
\end{deluxetable*}

\begin{figure*}[!t]
\centering
\includegraphics[width=0.95\textwidth]{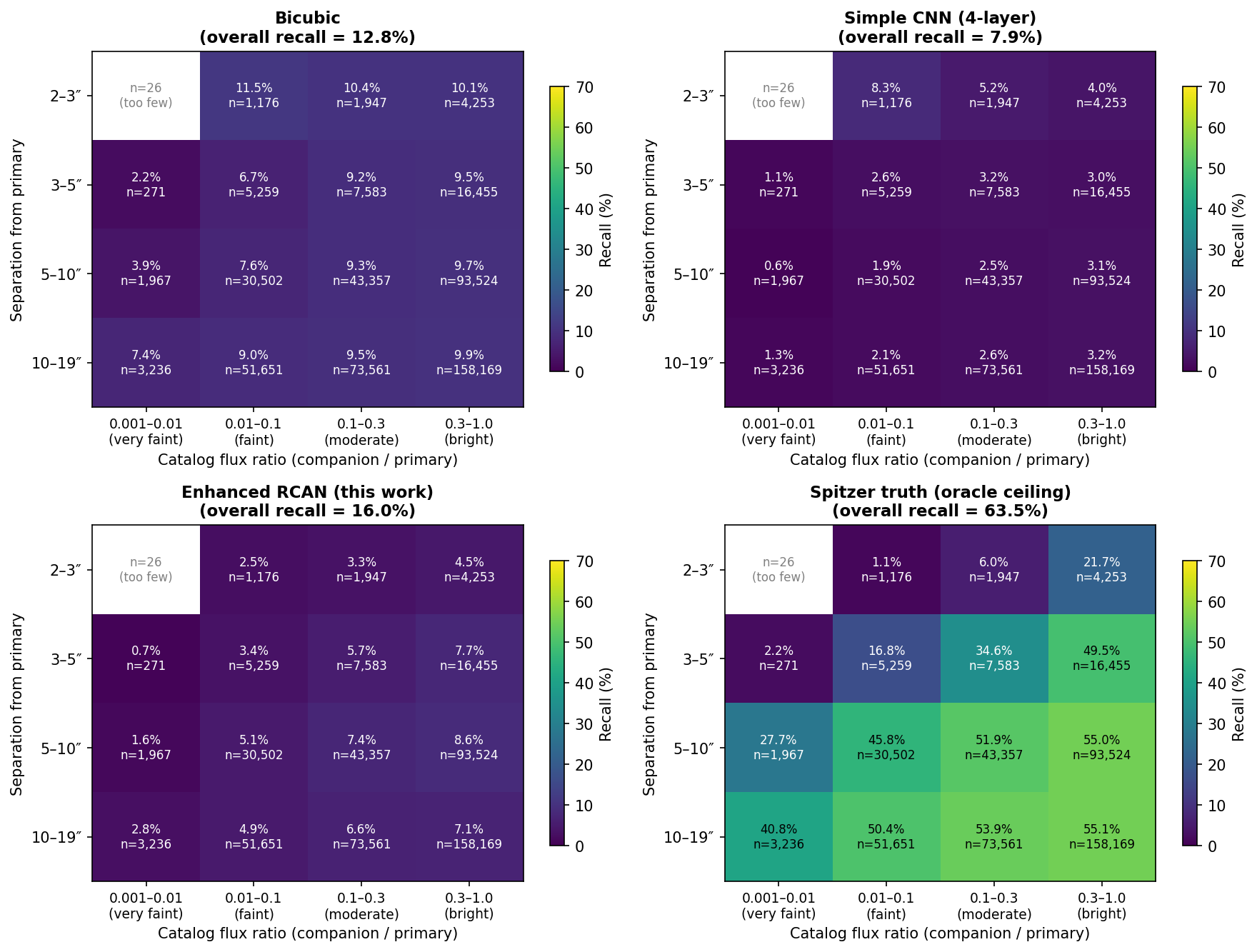}
\caption{Faint companion detection recall versus (separation from primary, catalog flux ratio) for each method. The Enhanced RCAN (bottom-left) substantially exceeds the interpolation baselines (top row) across the entire (separation, ratio) grid; the Spitzer truth oracle (bottom-right) sets the upper bound on what is detectable in this setup.}
\label{fig:companion_heatmap}
\end{figure*}

The Enhanced RCAN recovers $\approx 25\%$ of catalog companions detectable in the actual Spitzer truth (Table~\ref{tab:companion}), a factor of $\approx 1.3$ better than the strongest interpolation baseline and $\approx 2\times$ better than the trained Simple CNN baseline. The 2D recall structure (Figure~\ref{fig:companion_heatmap}) shows monotonic improvement with both separation and flux ratio for our model. Simple CNN underperforms even bicubic on faint companion detection: its oversmoothing destroys the fine peak structure required to detect dim sources, consistent with the failure mode analysis in Section~\ref{sec:failure_mode}.

\paragraph{Caveats.} The catalog flux is measured in IRAC Ch1 in deeper imaging than the per-cutout Spitzer truth used here; some catalog sources lie below the per-cutout local background in the actual Spitzer mosaic and are fundamentally undetectable in our setup. The Spitzer ceiling normalization above accounts for this effect. The position projection assumes a standard astronomical WCS convention (RA increases right-to-left, Dec increases upward), sanity-checked by confirming that the primary catalog source consistently sits near the cutout center. We caution that the peak detection threshold of 0.3 is a methodological choice; lower thresholds raise the apparent recall of all methods at the cost of a higher false-positive rate, but the qualitative finding is robust.

\subsection{Failure Mode: Oversmoothing of Source Profiles}
\label{sec:failure_mode}

\begin{figure*}[!t]
\centering
\includegraphics[width=0.95\textwidth]{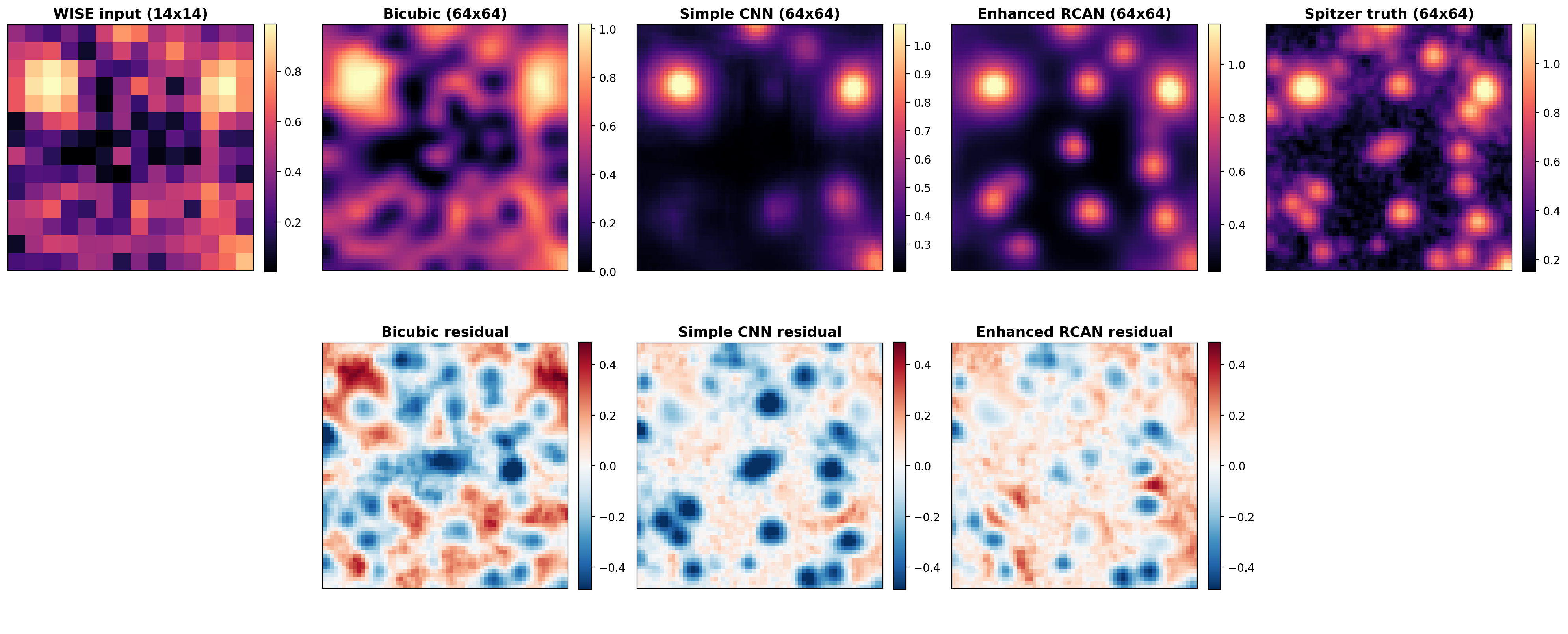}
\caption{Enhanced RCAN reconstruction of a representative bright source cutout, compared with the baseline methods. \textit{Top row:} the WISE input, the three super-resolution methods, and the Spitzer ground truth. The Spitzer truth resolves multiple distinct sources across the field, and the Enhanced RCAN output preserves the same sources with comparable sharpness. The bicubic upsampling smooths every source into a broad feature with no clear separation between neighbors, and the Simple CNN retains only the two brightest sources and flattens the rest into the background. \textit{Bottom row:} residual maps (method minus Spitzer truth) on a symmetric diverging scale. Blue marks a flux deficit at source cores and red marks excess flux in the surrounding wings. The Enhanced RCAN residuals are the smallest in amplitude and the most tightly confined to source positions. The remaining pattern of a core deficit surrounded by excess flux is the signature of oversmoothing, the characteristic failure mode shared by all methods and discussed in the text.}
\label{fig:failure_mode}
\end{figure*}

The dominant failure pattern across all super-resolution methods is oversmoothing of source profiles, visible directly in the image panels of Figure~\ref{fig:failure_mode} and quantified by the residual maps in its bottom row. All methods broaden the reconstructed sources relative to the Spitzer ground truth, spreading flux over a wider area than the compact Spitzer PSF. In the residual maps this appears as a characteristic redistribution of flux: a deficit at each source core, ringed by an excess in the surrounding wings. The effect is severe for bicubic and Simple CNN, which render sources as smooth Gaussian-like features with little of the compact structure of the Spitzer PSF and show the strongest flux redistribution; bicubic in particular merges neighboring sources into ill-defined envelopes. The Enhanced RCAN reproduces the Spitzer source morphology much more accurately, with residuals that are smaller in amplitude and more confined to source cores, but still oversmooths the wings slightly. This pattern is consistent with training objectives that minimize per-pixel error: smoothing toward the local mean is a low-loss prediction in the absence of explicit priors on the source profile \citep[e.g.,][]{Ledig2017, Zhao2017}.

This characteristic failure produces low per-pixel error (good MAE/RMSE/SSIM in the central few pixels) but disproportionately high integrated flux error, because the oversmoothing redistributes flux across many pixels and the cumulative bias compounds. The pattern is most pronounced for the faintest bin, where the signal-to-noise is lowest (Section~\ref{sec:brightness} shows the RCAN aperture integrated flux error to be $13\%$ at faint vs $8\%$ at bright).

This failure mode plausibly biases faint substructure and low surface brightness features, so morphological studies built on these outputs will need validation on a per-galaxy basis.

\subsection{Comparison to Baseline Methods}
\label{sec:comparison}

Table~\ref{tab:comparison} compares the Enhanced RCAN to the three baselines on the held-out test set.

\begin{deluxetable*}{lccccccccc}
\tablecaption{Method Comparison on the Test Set\label{tab:comparison}}
\tablewidth{0pt}
\tablehead{
    \colhead{Method} & \colhead{Params\tablenotemark{a}} & \colhead{SSIM\tablenotemark{b}} & \colhead{MAE\tablenotemark{c}} & \colhead{RMSE\tablenotemark{c}} & \colhead{Pearson $r$\tablenotemark{d}} & \colhead{Image PSNR\tablenotemark{b}} & \colhead{Peak err\tablenotemark{e}} & \colhead{Int.\ flux err\tablenotemark{f}} & \colhead{SNR pres.\tablenotemark{g}}
}
\startdata
Bicubic & --- & $0.29$ & $0.160$ & $0.199$ & $0.59$ & $20.19$ dB & $20.3\%$ & $18.2\%$ & $1.04$ \\
Bilinear & --- & $0.33$ & $0.149$ & $0.185$ & $0.61$ & $20.83$ dB & $23.0\%$ & $20.0\%$ & $1.04$ \\
Simple CNN (4 layers) & 67k & $0.45$ & $0.109$ & $0.145$ & $0.71$ & $23.01$ dB & $15.6\%$ & $30.4\%$ & $0.86$ \\
Enhanced RCAN (this work) & 12.5M & 0.51 & 0.100 & 0.137 & 0.75 & 23.6 dB & 7.6\% & 20.4\% & $0.97$ \\
\enddata
\tablenotetext{a}{Trainable parameter count.}
\tablenotetext{b}{Mean over per-cutout values at $\mathrm{max\_val} = 2.0$.}
\tablenotetext{c}{Aggregated over all pixels in 83,592 test cutouts in normalized space.}
\tablenotetext{d}{Per-pixel correlation between predicted and Spitzer truth output.}
\tablenotetext{e}{Mean per-image relative error in peak flux (brightest pixel anywhere in cutout).}
\tablenotetext{f}{Mean per-image relative error in integrated flux at peak threshold $\tau = 0.5$.}
\tablenotetext{g}{mean(signal)/std(background); see text.}
\end{deluxetable*}

Deeper architectures progressively improve most metrics, but two findings deserve further consideration. First, the Simple CNN integrated flux error ($30.4\%$) is worse than bicubic's ($18.2\%$) despite Simple CNN being lower on every other metric. Shallow CNNs trained on mean squared error oversmooth source pixels heavily, lowering per-pixel error while degrading the cumulative integrated flux quantity (Section~\ref{sec:failure_mode}). The deeper architecture and source-focused loss of Enhanced RCAN recover this aspect of the photometry. Second, the ``SNR preservation'' column is the least discriminating of the metrics: the interpolation baselines ($1.04$) and Enhanced RCAN ($0.97$) both sit near the nominal target of unity. We caution that proximity to unity here does not indicate reconstruction quality: the metric measures whether marginal source/background statistics are preserved, not whether structure is recovered. Interpolation preserves marginal statistics nearly by construction while reconstructing little of the underlying structure (low SSIM, high peak flux error), whereas Enhanced RCAN reaches comparable preservation of marginal statistics together with substantially better structural and photometric reconstruction.

\subsection{An End-to-end Example}
\label{sec:example}

\begin{figure*}[!t]
\centering
\includegraphics[width=\textwidth]{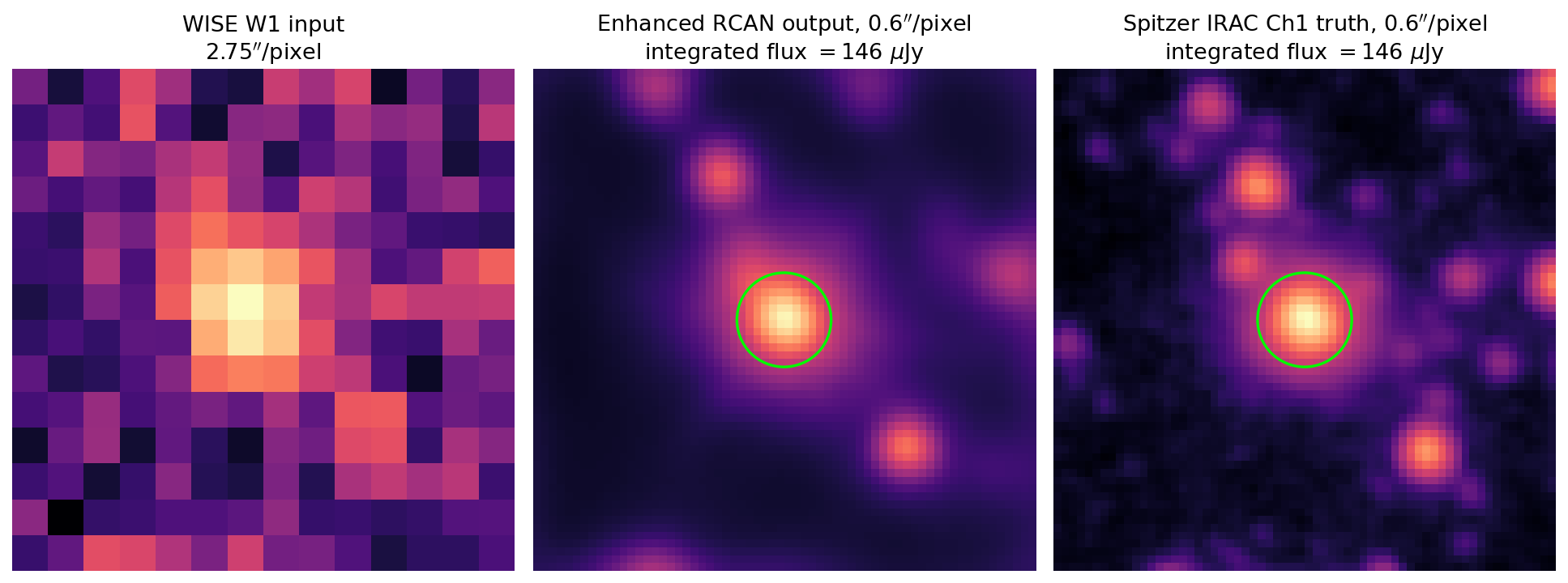}
\caption{End-to-end application to a single COSMOS2020 galaxy (ID 376723, $z = 0.34$). Left: the \wise W1 input cutout. Middle: the Enhanced RCAN output. Right: the \spitzer \irac Ch1 ground truth. The green circle marks the integrated aperture ($r = 3.6\arcsec$) at the catalog position. Fluxes are measured after inverting the normalization to physical units and subtracting the local sky in an annulus. The model recovers the integrated aperture flux of the Spitzer truth to better than $1\%$ for this galaxy, and the neighboring sources in the output correspond to real sources in the truth. The aperture peak flux is $7\%$ low, consistent with the mild oversmoothing of Section~\ref{sec:failure_mode}.}
\label{fig:galaxy_case}
\end{figure*}

Figure~\ref{fig:galaxy_case} shows the intended use case executed end to end on a single galaxy. The object is COSMOS2020 source 376723, a galaxy at photometric redshift $z = 0.34$ with a catalog Ch1 flux of $177\ \mu$Jy \citep{Weaver2022}. The \wise input shows a single blended feature. The model output resolves the galaxy and its neighboring sources at Spitzer resolution. We invert the normalization to physical units and apply the fixed apertures of Section~\ref{sec:photometry}. The same local sky annulus is subtracted from both images, so the comparison remains differential. The integrated aperture flux of the model output agrees with that of the Spitzer truth to $0.4\%$ ($146\ \mu$Jy in both). Without the sky subtraction the agreement is $0.7\%$. The fixed $3.6\arcsec$ aperture captures $83\%$ of the catalog total flux in the truth image, as expected for an extended source. This example is illustrative rather than typical. The population median integrated flux error is $10.8\%$ (Section~\ref{sec:photometry}), and single object performance varies accordingly.

\subsection{Deployment Beyond the Training Field}
\label{sec:extrapolation}

\begin{figure*}[!t]
\centering
\includegraphics[width=0.78\textwidth]{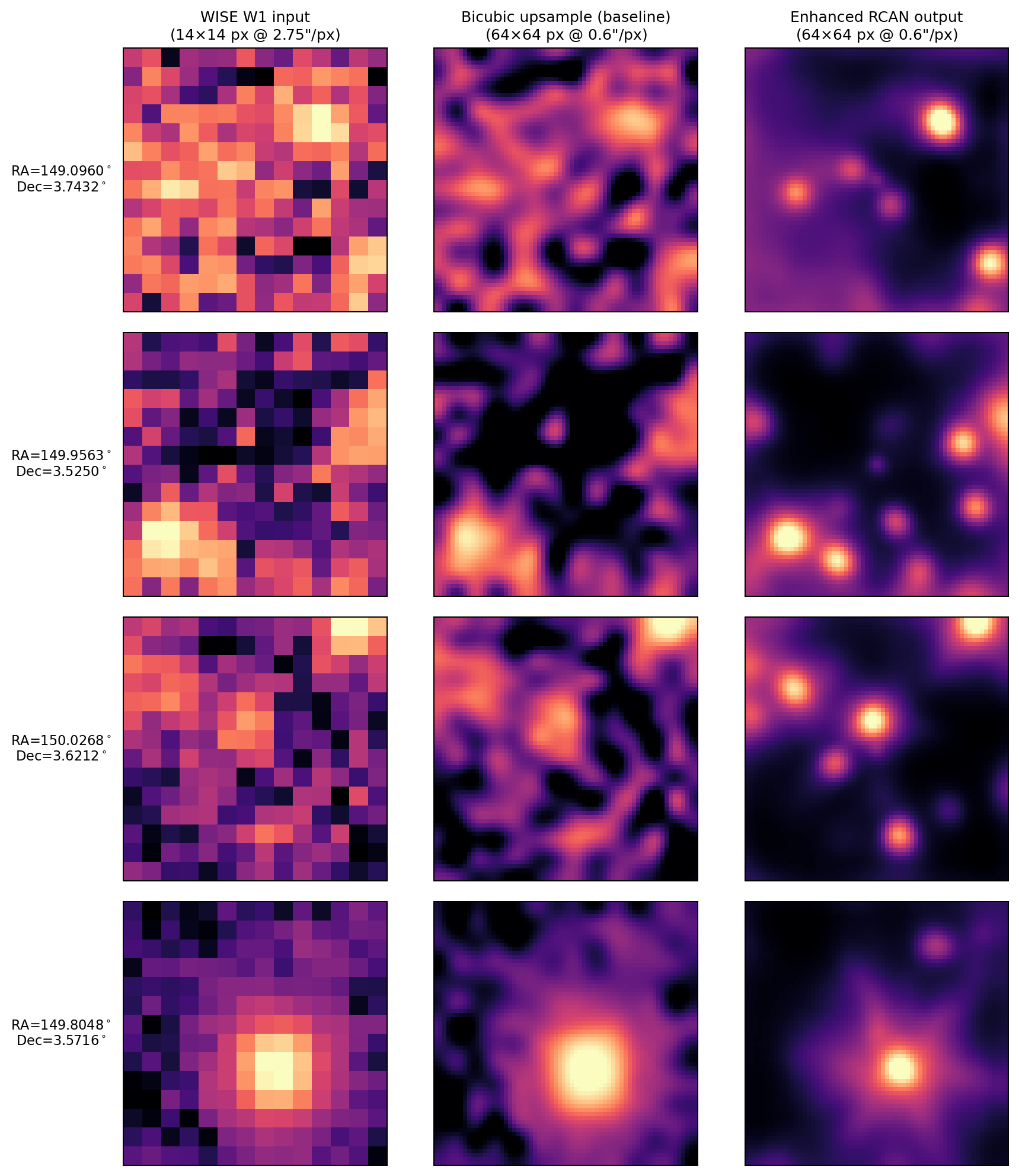}
\caption{Comparison of \wise input (left), bicubic upsampling (middle), and Enhanced RCAN output (right) for four sky positions inside \unwise tile coverage but outside the \spitzer-\cosmos mosaic. No \spitzer ground truth is available for any of these positions. Rows are ordered by increasing input intensity. The Enhanced RCAN output is qualitatively consistent with the model's output inside the COSMOS field, suggesting that the learned \wise$\to$~\spitzer mapping is not specific to COSMOS-field source density patterns. Quantitative validation against external Spitzer or JWST imaging where it exists is left to future work.}
\label{fig:extrapolation_demo}
\end{figure*}

The headline implication of the characterization above is that the trained model can be applied to \wise data outside the COSMOS field, where no \spitzer ground truth is available. To check this empirically, we run the model on \wise W1 cutouts drawn from sky positions that lie inside the \unwise tile coverage but outside the \spitzer-\cosmos mosaic footprint, where the model has neither training data nor any per-cutout ground truth (Figure~\ref{fig:extrapolation_demo}). The model produces qualitatively well-behaved output: source-like peaks where the \wise input shows excess flux, smooth backgrounds elsewhere, and no obvious COSMOS-specific artifacts. We caution that this is a generalization sanity check, not a quantitative claim, but it is the result that justifies the practical case for the model: \wise has all-sky coverage that \spitzer never achieved, and the purpose of the model is to enhance \wise data wherever \wise data exist.

\subsection{Domain of Validity}
\label{sec:limitations}

The characterization above defines the domain in which the model can be trusted, and we collect those boundaries here.

The model is trained exclusively on the \cosmos field, and the high source density of our test set (Section~\ref{sec:cosmos}) makes the dataset unrepresentative of less-crowded sky regions, so performance on Galactic-plane fields, galaxy clusters, or sparser regions may differ. The training set is likewise dominated by compact sources, so very extended emission (diffuse ISM, galaxy halos, nearby galaxies) is underrepresented, and sources approaching detector saturation in either \wise or \spitzer are unlikely to be accurately reconstructed. We also assume negligible color differences between \wise W1 (3.4\,\micron) and \spitzer Ch1 (3.6\,\micron), and for sources with unusual spectral energy distributions (AGN, high-redshift galaxies) color-dependent residuals may arise.

Two boundaries concern our measurements rather than the model. The threshold-based source definitions used in the threshold mask metrics (Sections~\ref{sec:photometry} and \ref{sec:deblending}) are heuristic rather than astronomically rigorous. We mitigate this by reporting metrics based on aperture photometry (Table~\ref{tab:photometry_aperture}) and by using catalog-based ground truth for companion detection (Section~\ref{sec:companion}), and the qualitative findings are consistent across all definitions, though neither definition is fully ideal in this dense-field regime. Absolute photometry in fields as dense as COSMOS also requires a background method suited to crowding rather than a sky annulus, for the reasons set out in Section~\ref{sec:photometry}.

Several of these boundaries are addressable rather than fundamental, and we return to them in the directions for future work in Section~\ref{sec:conclusions}.

\section{DISCUSSION}
\label{sec:discussion}

\subsection{Physical Interpretation}
\label{sec:physical}

The success of the super-resolution approach relies on the statistical relationship between the two instruments' imaging, which the network learns from paired examples. The framework itself is not specific to \wise and \spitzer. It requires only two instruments with overlapping coverage, a resolution gap, and enough paired cutouts to train on, and Appendix~\ref{sec:appendix_ch2} demonstrates this by replicating the entire pipeline at W2 $\to$ \irac Ch2. The network learns three coupled transformations:

\begin{enumerate}
    \item \textbf{PSF deconvolution}: The dominant effect is the sharpening of the broader \wise PSF ($6.1\arcsec$ FWHM) toward the \spitzer PSF ($1.7\arcsec$ FWHM). This is an ill-posed inverse problem that the network solves using learned priors derived from the training data. The image panels and residual maps of Figure~\ref{fig:failure_mode} show that the reconstruction approaches but does not fully recover the Spitzer PSF compactness. We quantify this with a concentration index, the ratio of the aperture flux at $r = 2$ px to that at $r = 6$ px at the central catalog position. The Spitzer truth cutouts define the expected concentration at the Spitzer PSF. On the brightest quartile, where the truth is most compact, the RCAN output recovers a median $94\%$ of the expected concentration, versus $72\%$ for bicubic upsampling. On the faintest quartile the model output is instead more concentrated than the truth, with a median ratio of $1.26$. This matches the faint source behavior discussed in Section~\ref{sec:brightness}, where the model reconstructs sources sharply at the cost of amplifying noise.

    \item \textbf{Flux calibration}: The network absorbs residual calibration offsets between \wise and \spitzer that survive our unit conversion. After unit conversion, the median per-pixel surface brightness of the \wise cutouts remains a factor of $\sim 2.5$ higher than that of the matched \spitzer cutouts; the network learns this offset as part of the super-resolution mapping.

    \item \textbf{Noise characteristics}: \wise and \spitzer have different noise properties owing to differences in detector technology and observation strategy. The network learns to suppress \wise noise patterns while preserving genuine source structure.
\end{enumerate}

The combination of a factor of $\approx 1.5$--$2$ better aperture photometry accuracy than the interpolation baselines (Section~\ref{sec:photometry}) and $\approx 4\times$ better peak recovery at 3--5\arcsec\ separations (Section~\ref{sec:deblending}) suggests that the network performs genuine resolution enhancement and source deblending, rather than smoothing or artificial sharpening of the \wise input.

\subsection{Relation to Previous Super-resolution Efforts}
\label{sec:previous_work}

Deep-learning super-resolution has been demonstrated in several regimes. \citet{Schawinski2017} trained a generative adversarial network on SDSS galaxies to recover images artificially degraded in seeing and noise, quantified by PSNR alone. Their preprocessing clips extreme pixel values, which they note prevents recovery of the original flux calibration. \citet{Sweere2022} trained on simulated XMM-Newton images for a factor of 2 PSF reduction. They removed the adversarial discriminator because hallucinated features are unacceptable in astronomical imaging, and validated with image quality metrics and qualitative comparison to Chandra observations. Closest to our setting, \citet{Lauritsen2021} super-resolved Herschel SPIRE 500\,\micron\ imaging (PSF FWHM $36.6\arcsec$) toward JCMT SCUBA-2 450\,\micron\ ($7.9\arcsec$) on the same COSMOS field. They report completeness above $95\%$ for sources brighter than 15 mJy, purity above $87\%$, and an approximately one-to-one bright source flux relation, with completeness falling steeply toward fainter fluxes. Our findings match the patterns shared across these works. Fidelity degrades toward faint sources. Reconstructions inherit the training set prior, so rare source types are not guaranteed to be recovered. Avoiding adversarial training in favor of an explicitly source-focused objective trades some sharpness for reliability. Relative to these efforts, this work adds per-source aperture photometry in physical units across the full test set, brightness-resolved error and dispersion budgets, and a direct measurement of PSF compactness recovery.

\section{CONCLUSIONS}
\label{sec:conclusions}

In this paper, we have presented and characterized a deep-learning framework for enhancing \wise 3.4\,\micron\ imaging toward \spitzer 3.6\,\micron\ resolution. Our main findings are summarized below.

\begin{enumerate}
    \item Using a sample of $\sim 390{,}000$ paired \wise/\spitzer cutouts drawn uniformly from the \cosmos field, we have trained and characterized an Enhanced Residual Channel Attention Network (12.5\,M parameters) that performs $4.6\times$ spatial super-resolution from \wise W1 ($2.75\arcsec$ pixel$^{-1}$) to \spitzer \irac Ch1 ($0.6\arcsec$ pixel$^{-1}$) sampling.

    \item The image quality is moderate: SSIM\,$=$\,0.51 (at $\mathrm{max\_val} = 2.0$), image PSNR\,$\approx$\,23.6\,dB, and Pearson $r$\,$=$\,0.75.

    \item Using aperture photometry at the central catalog source, we find that the model recovers peak flux to a median relative error of $20\%$ and integrated flux to $11\%$, both a factor of $\approx 1.5$--$2$ better than the interpolation baselines. The lead grows strongly with source brightness: the aperture peak error drops to $13\%$ on the brightest quartile of test sources, while the interpolation baselines remain at $\approx 25$--$50\%$ across the bins (Section~\ref{sec:brightness}).

    \item At 3--5\arcsec\ separations where \wise\ blends sources that \spitzer\ separates, the model recovers $35\%$ of the source peaks detectable in the Spitzer truth compared with $9\%$ for bicubic upsampling, a factor of $\approx 3.8$ improvement.

    \item For faint companion detection against COSMOS catalog ground truth, the model recovers $\approx 25\%$ of catalog companions detectable in the actual Spitzer cutouts (vs $\approx 20\%$ for bicubic; Spitzer ceiling 63\%).

    \item The characteristic failure mode is oversmoothing of source profiles, which produces low per-pixel error but compounds into an integrated flux bias that is most severe for faint sources. The pattern is shared with all interpolation baselines but is much less pronounced for the trained model.

\end{enumerate}

The conclusion of the \neowise mission has made methods that maximize the scientific return from existing \wise data increasingly valuable. The characterization above suggests that the model is most useful for faint and intermediate brightness sources, where the integrated flux gain over interpolation reaches a factor of $\approx 3$. The model also retains the lowest integrated flux error on the brightest quartile. Suitability for specific downstream applications (galaxy morphology, source deblending in particular target fields, faint companion follow-up) requires study-specific validation that this paper does not provide; the deblending and companion detection capabilities reported here are intended as a starting point.

The trained model and preprocessing code are publicly available at \url{https://github.com/srezaeeucr/wiser-sr}. Several directions for future work suggest themselves. First, retraining with explicit source profile or perceptual loss terms may mitigate the oversmoothing failure mode. Second, the framework can be extended to additional \wise bands and multiband super-resolution. Third, the model's performance should be validated across diverse sky environments outside the \cosmos field, where deeper external imaging exists.

\let\internallinenumbers\relax
\begin{acknowledgments}
This publication makes use of data products from the Wide-field Infrared Survey Explorer, which is a joint project of the University of California, Los Angeles, and the Jet Propulsion Laboratory/California Institute of Technology, funded by the National Aeronautics and Space Administration.

This work is based in part on observations made with the Spitzer Space Telescope, which was operated by the Jet Propulsion Laboratory, California Institute of Technology under a contract with NASA.
\end{acknowledgments}

\facilities{WISE, Spitzer (IRAC)}

\software{TensorFlow \citep{tensorflow2015}, NumPy \citep{numpy2020}, Astropy \citep{astropy2022}, Matplotlib \citep{matplotlib2007}, scikit-image \citep{scikitimage2014}}

\FloatBarrier  
\appendix

\section{ARCHITECTURE, LOSS, AND TRAINING DETAILS}
\label{appendix:methods}

This appendix gives the full technical specification of the network, the loss function, and the training configuration summarized in Sections~\ref{sec:architecture}--\ref{sec:training}.

\subsection{Multiscale Input Block}
\label{sec:multiscale}

Rather than processing the input at a single scale, we extract features at three scales simultaneously. The first branch applies a 3$\times$3 convolution at the original scale. The second upsamples the input by a factor of 2, applies a 5$\times$5 convolution, and downsamples back. The third applies a 5$\times$5 convolution at the original scale, giving a larger receptive field. The three feature sets are concatenated and fused via a $1 \times 1$ convolution, supplying the network with information at multiple spatial frequencies from the outset.

\subsection{Residual Channel Attention Blocks}

The structure inherited from \citet{Zhang2018RCAN} is residual-in-residual: residual groups containing residual channel attention blocks, with long and short skip connections at multiple levels. The channel attention module follows the squeeze-and-excitation design of \citet{Hu2018SENet}, with global average pooling and reduction ratio 16.

The core of the network consists of Residual Channel Attention Blocks (RCABs), which combine residual learning with channel-wise attention. Each RCAB comprises two 3$\times$3 convolutional layers with ReLU activation, followed by a squeeze-and-excitation channel attention module, with a residual connection scaled by a factor of 0.1.

The channel attention module adaptively recalibrates channel-wise feature responses:
\begin{equation}
    \mathbf{y} = \mathbf{x} \cdot \sigma\left(W_2 \cdot \mathrm{ReLU}\left(W_1 \cdot \mathrm{GAP}(\mathbf{x})\right)\right)
\end{equation}
where GAP denotes global average pooling, $W_1$ and $W_2$ are fully connected layers with reduction ratio 16, and $\sigma$ is the sigmoid function.

The RCABs are organized into 4 residual groups of 8 blocks each (32 in total), with group-level skip connections that enable deep feature extraction while preserving stable gradient flow.

\subsection{Progressive Upsampling}
\label{sec:upsampling}

The spatial resolution is increased from $14 \times 14$ to $64 \times 64$ through three successive 2$\times$ subpixel convolutions, taking the features to $28 \times 28$, $56 \times 56$, and $112 \times 112$, followed by a center crop to $64 \times 64$.

Subpixel convolution \citep[pixel shuffle;][]{Shi2016} rearranges features from the channel dimension to the spatial dimensions, providing a learned upsampling that outperforms interpolation. The final crop from $112 \times 112$ to $64 \times 64$ avoids the edge artifacts that would otherwise arise from padding-based upsampling. Intermediate outputs at $28 \times 28$ and $56 \times 56$ are tap-offs for progressive supervision during training. The intermediate gradient signal stabilizes training across the $4.6\times$ upsampling factor.

\subsection{Loss Equations}
\label{appendix:loss_equations}

Source pixels are those above a threshold $\tau$, set to 0.5 in normalized units (approximately the median source flux). The weighted Huber loss takes the form
\begin{equation}
    \mathcal{L}_\mathrm{Huber} = \frac{1}{N}\sum_i w_i \cdot H_\delta(y_i - \hat{y}_i)
\end{equation}
where the Huber function $H_\delta$ provides L2 behavior for small errors and L1 for large errors:
\begin{equation}
    H_\delta(e) = \begin{cases}
        \frac{1}{2}e^2 & |e| \leq \delta \\
        \delta(|e| - \frac{1}{2}\delta) & |e| > \delta
    \end{cases}
\end{equation}
with $\delta = 1.0$, and the per-pixel weights are:
\begin{equation}
    w_i = \begin{cases}
        w_\mathrm{source} = 3.0 & y_i > \tau \\
        w_\mathrm{background} = 0.5 & y_i \leq \tau
    \end{cases}
\end{equation}

The Structural Similarity Index \citep[SSIM;][]{Wang2004} quantifies perceptual image quality by comparing luminance, contrast, and structure. The SSIM loss term is
\begin{equation}
    \mathcal{L}_\mathrm{SSIM} = 1 - \mathrm{SSIM}(y, \hat{y})
\end{equation}

To preserve sharp edges around sources, the spatial-gradient loss term is
\begin{equation}
    \mathcal{L}_\mathrm{grad} = |\nabla_x y - \nabla_x \hat{y}| + |\nabla_y y - \nabla_y \hat{y}|
\end{equation}
where $\nabla_x$ and $\nabla_y$ denote spatial gradients.

The total loss combines the three components with progressive supervision at the three output scales:
\begin{equation}
    \mathcal{L}_\mathrm{total} = \sum_{s \in \{28, 56, 64\}} \lambda_s \left( \alpha \mathcal{L}_\mathrm{Huber}^s + \beta \mathcal{L}_\mathrm{SSIM}^s + \gamma \mathcal{L}_\mathrm{grad}^s \right)
\end{equation}
with $\lambda_{28} = 0.2$, $\lambda_{56} = 0.3$, $\lambda_{64} = 0.5$, and $\alpha = 0.5$, $\beta = 0.35$, $\gamma = 0.15$. We set these values by hand during model development. The scale weights $\lambda_s$ increase toward the final $64 \times 64$ output because that is the resolution the model is evaluated at; the intermediate heads serve to stabilize training. The component weights rank pixel fidelity first, structural similarity second, and the gradient term as a small regularizer. Candidate settings were compared by validation performance across successive training runs; we did not perform an exhaustive search over these weights.

\subsection{Training Configuration}
\label{appendix:training_config}

We use an initial learning rate of $2 \times 10^{-4}$, weight decay of $5 \times 10^{-4}$, and batch size 32. The learning rate decays exponentially at a rate of 0.98 per 1,000 steps, and training stops early when the validation loss has not improved for 40 epochs.

Training is performed with mixed-precision (float16) computation, using TensorFlow 2.x on NVIDIA GPU hardware. Convergence typically occurs within 30 epochs and requires $\approx 12$ hours on a single GPU. We do not apply data augmentation; the large training set ($\sim 390{,}000$ samples) provides sufficient diversity without it.

\section{REPLICATION ON \wise{} W2 / \spitzer{} \irac{} Ch2}
\label{sec:appendix_ch2}

We replicated the entire pipeline at the next-bluest pair of channels, \wise W2 (4.6\,\micron) $\to$ \spitzer \irac Ch2 (4.5\,\micron), to test whether the framework's behavior is specific to the W1\,/\,Ch1 bandpass or generalizes across the available infrared channel pairs. The replication used the same pipeline as Sections~\ref{sec:methods}--\ref{sec:results} throughout, with identical extraction logic and preprocessing, the same Enhanced RCAN architecture, the same source-focused composite loss, and the same optimizer schedule. Its training pool was constructed identically to that of the W1 $\to$ Ch1 reference model, giving 390{,}170 samples from the same 70/15/15 stratified split applied to the Ch2 extraction. Only the input WISE tile band and the target Spitzer mosaic were swapped.

Tables~\ref{tab:appendix_aperture}, \ref{tab:appendix_deblending}, and~\ref{tab:appendix_companion} report the headline aperture photometry, deblending, and companion detection metrics side by side for the two channels. RCAN's median aperture relative error on Ch2 is lower than on Ch1 across both the peak (2 px radius: 0.184 vs 0.201) and integrated (6 px radius: 0.097 vs 0.108) apertures, suggesting that the W2 $\to$ Ch2 mapping is no harder to learn than W1 $\to$ Ch1 despite W2's slightly larger PSF (FWHM\,$\approx 6.4\arcsec$ versus W1's $6.1\arcsec$). The deblending recall on Ch2 (0.32) is below that on Ch1 (0.40), consistent with Ch2's lower native source detection signal-to-noise ratio at fixed catalog completeness, but the multiplicative improvement over bicubic upsampling is preserved (3.96$\times$ on Ch2 versus 3.82$\times$ on Ch1). Faint companion recovery follows the same pattern: the absolute recall is lower on Ch2 (0.152 versus 0.160), but the ratio over bicubic (1.25$\times$ Ch2 versus 1.25$\times$ Ch1) is identical to within rounding. The dispersion and compactness measures replicate as well. The RCAN integrated aperture NMAD is $14.4\%$ on Ch2 versus $15.9\%$ on Ch1, the smallest of the four methods in both channels. On the brightest quartile the RCAN output recovers $94\%$ of the expected Spitzer PSF concentration in both channels.

\begin{deluxetable*}{llcc}
\tablecaption{Aperture Photometry Accuracy: Ch1 versus Ch2\label{tab:appendix_aperture}}
\tablewidth{0pt}
\tablehead{
    \colhead{Aperture} & \colhead{Method} & \colhead{Ch1\tablenotemark{a}} & \colhead{Ch2\tablenotemark{b}}
}
\startdata
Peak ($r = 2$ px) & Bicubic    & 0.320 & 0.319 \\
        & Bilinear   & 0.302 & 0.294 \\
        & Simple CNN & 0.271 & 0.241 \\
        & RCAN & 0.201 & 0.184 \\
\hline
Integrated ($r = 6$ px) & Bicubic    & 0.225 & 0.210 \\
        & Bilinear   & 0.217 & 0.199 \\
        & Simple CNN & 0.168 & 0.148 \\
        & RCAN & 0.108 & 0.097 \\
\enddata
\tablenotetext{a}{Median relative error in aperture flux against Spitzer ground truth on the W1\,$\to$\,Ch1 test set ($n = 83{,}592$). Lower values indicate higher accuracy.}
\tablenotetext{b}{Same on the W2\,$\to$\,Ch2 test set ($n = 83{,}609$); Ch2 values are uniformly below Ch1, indicating that W2's slightly larger PSF does not impair the framework.}
\end{deluxetable*}

\begin{deluxetable*}{lcccc}
\tablecaption{Source-Deblending Performance at Peak Threshold $= 1.0$\label{tab:appendix_deblending}}
\tablewidth{0pt}
\tablehead{
    \colhead{Method} & \colhead{Ch1 recall\tablenotemark{a}} & \colhead{Ch2 recall\tablenotemark{a}} & \colhead{Ch1 precision\tablenotemark{b}} & \colhead{Ch2 precision\tablenotemark{b}}
}
\startdata
Bicubic     & 0.103 & 0.081 & 0.596 & 0.572 \\
Bilinear    & 0.076 & 0.060 & 0.522 & 0.509 \\
Simple CNN  & 0.245 & 0.167 & 0.754 & 0.713 \\
RCAN        & 0.395 & 0.321 & 0.692 & 0.640 \\
\hline
RCAN/bicubic recall ratio\tablenotemark{c} & 3.82$\times$ & 3.96$\times$ & --- & --- \\
\enddata
\tablenotetext{a}{Fraction of Spitzer truth peaks within 1.5\arcsec\ of a peak detected by the method, evaluated over the full test set (all separations). Same evaluator code as Section~\ref{sec:results}.}
\tablenotetext{b}{Fraction of peaks detected by each method that match a Spitzer truth peak.}
\tablenotetext{c}{Multiplicative improvement of RCAN recall over bicubic recall in each band; the ratio is preserved across bands and is slightly larger for Ch2.}
\end{deluxetable*}

\begin{deluxetable*}{lccc}
\tablecaption{Faint Companion Recovery Against Catalog Truth\label{tab:appendix_companion}}
\tablewidth{0pt}
\tablehead{
    \colhead{Method} & \colhead{Ch1 recall\tablenotemark{a}} & \colhead{Ch2 recall\tablenotemark{b}} & \colhead{Ch2/Ch1\tablenotemark{c}}
}
\startdata
Spitzer truth (ceiling) & 0.635 & 0.620 & 0.98 \\
Bicubic    & 0.128 & 0.122 & 0.95 \\
Bilinear   & 0.120 & 0.116 & 0.97 \\
Simple CNN & 0.079 & 0.072 & 0.91 \\
RCAN       & 0.160 & 0.152 & 0.95 \\
\enddata
\tablenotetext{a}{Fraction of COSMOS2020 catalog companions whose projected pixel position lies within 1.5\arcsec\ of a peak in the method's W1\,$\to$\,Ch1 output. Catalog companions: 1{,}029{,}115 across 83{,}592 test cutouts.}
\tablenotetext{b}{Same on the W2\,$\to$\,Ch2 test set; catalog companions 1{,}021{,}878 across 83{,}609 cutouts.}
\tablenotetext{c}{Ch2 recall divided by Ch1 recall; all five rows including the truth ceiling are $\sim 0.95$, indicating that the gap is set by Ch2's intrinsic signal-to-noise rather than by the model.}
\end{deluxetable*}

The Ch2 replication confirms that the framework's behavior is band-agnostic across the W1/W2 $\to$ Ch1/Ch2 channel pairs at the level of $\sim 10\%$ in absolute metric values. We caution that further extension to the W3 (12\,\micron) and W4 (22\,\micron) channels would require separate validation and is beyond the scope of this paper. Their PSFs are much broader, and they target IRAC Ch3/Ch4 at higher resolution ratios and different astrophysical regimes.

\bibliography{references}

\begin{thebibliography}{}
\expandafter\ifx\csname natexlab\endcsname\relax\def\natexlab#1{#1}\fi
\providecommand{\url}[1]{\href{#1}{#1}}
\providecommand{\dodoi}[1]{doi:~\href{http://doi.org/#1}{\nolinkurl{#1}}}
\providecommand{\doeprint}[1]{\href{http://ascl.net/#1}{\nolinkurl{http://ascl.net/#1}}}
\providecommand{\doarXiv}[1]{\href{https://arxiv.org/abs/#1}{\nolinkurl{https://arxiv.org/abs/#1}}}

\bibitem[{{Abadi} {et~al.}(2015)}]{tensorflow2015}
{Abadi}, M., {et~al.} 2015, {TensorFlow}: Large-Scale Machine Learning on
  Heterogeneous Systems, \url{https://www.tensorflow.org/}

\bibitem[{{Astropy Collaboration}(2022)}]{astropy2022}
{Astropy Collaboration}. 2022, The Astrophysical Journal, 935, 167,
  \dodoi{10.3847/1538-4357/ac7c74}

\bibitem[{{Bertin} \& {Arnouts}(1996)}]{Bertin1996}
{Bertin}, E., \& {Arnouts}, S. 1996, \aaps, 117, 393,
  \dodoi{10.1051/aas:1996164}

\bibitem[{{Calzetti} {et~al.}(2000){Calzetti}, {Armus}, {Bohlin}, {Kinney},
  {Koornneef}, \& {Storchi-Bergmann}}]{Calzetti2000}
{Calzetti}, D., {Armus}, L., {Bohlin}, R.~C., {et~al.} 2000, \apj, 533, 682,
  \dodoi{10.1086/308692}

\bibitem[{{Casey} {et~al.}(2014){Casey}, {Narayanan}, \& {Cooray}}]{Casey2014}
{Casey}, C.~M., {Narayanan}, D., \& {Cooray}, A. 2014, Physics Reports, 541,
  45, \dodoi{10.1016/j.physrep.2014.02.009}

\bibitem[{{Chary} \& {Elbaz}(2001)}]{Chary2001}
{Chary}, R., \& {Elbaz}, D. 2001, \apj, 556, 562, \dodoi{10.1086/321609}

\bibitem[{{Cosmic Dawn Team}(2021)}]{CosmicDawn2021}
{Cosmic Dawn Team}. 2021, {Spitzer Cosmic Dawn Survey},  IPAC,
  \dodoi{10.26131/IRSA566}

\bibitem[{{COSMOS Team}(2022)}]{COSMOS2020classic2022}
{COSMOS Team}. 2022, {COSMOS2020 Classic Catalog},  IPAC,
  \dodoi{10.26131/IRSA563}

\bibitem[{{Cutri} {et~al.}(2013){Cutri}, {Wright}, {Conrow},
  {et~al.}}]{Cutri2013}
{Cutri}, R.~M., {Wright}, E.~L., {Conrow}, T., {et~al.} 2013, {Explanatory
  Supplement to the AllWISE Data Release Products}, IPAC/Caltech

\bibitem[{{Dabbech} {et~al.}(2022){Dabbech}, {Terris}, {Jackson}, {Ramatsoku},
  {Smirnov}, \& {Wiaux}}]{Dabbech2022}
{Dabbech}, A., {Terris}, M., {Jackson}, A., {et~al.} 2022, \apjl, 939, L4,
  \dodoi{10.3847/2041-8213/ac98af}

\bibitem[{{D{\'i}az Baso} \& {Asensio Ramos}(2018)}]{Diaz2018}
{D{\'i}az Baso}, C.~J., \& {Asensio Ramos}, A. 2018, Astronomy \& Astrophysics,
  614, A5, \dodoi{10.1051/0004-6361/201731344}

\bibitem[{{Dieleman} {et~al.}(2015){Dieleman}, {Willett}, \&
  {Dambre}}]{Dieleman2015}
{Dieleman}, S., {Willett}, K.~W., \& {Dambre}, J. 2015, \mnras, 450, 1441,
  \dodoi{10.1093/mnras/stv632}

\bibitem[{{Dong} {et~al.}(2014){Dong}, {Loy}, {He}, \& {Tang}}]{Dong2014}
{Dong}, C., {Loy}, C.~C., {He}, K., \& {Tang}, X. 2014, in {Proceedings of the
  European Conference on Computer Vision (ECCV)}, 184--199,
  \dodoi{10.1007/978-3-319-10593-2_13}

\bibitem[{{Draine} \& {Li}(2007)}]{Draine2007}
{Draine}, B.~T., \& {Li}, A. 2007, \apj, 657, 810, \dodoi{10.1086/511055}

\bibitem[{{Elbaz} {et~al.}(2011){Elbaz}, {Dickinson}, {Hwang},
  {et~al.}}]{Elbaz2011}
{Elbaz}, D., {Dickinson}, M., {Hwang}, H.~S., {et~al.} 2011, \aap, 533, A119,
  \dodoi{10.1051/0004-6361/201117239}

\bibitem[{{Fazio} {et~al.}(2004){Fazio}, {Hora}, {Allen}, {et~al.}}]{Fazio2004}
{Fazio}, G.~G., {Hora}, J.~L., {Allen}, L.~E., {et~al.} 2004, The Astrophysical
  Journal Supplement Series, 154, 10, \dodoi{10.1086/422843}

\bibitem[{{Fruchter} \& {Hook}(2002)}]{FruchterHook2002}
{Fruchter}, A.~S., \& {Hook}, R.~N. 2002, \pasp, 114, 144,
  \dodoi{10.1086/338393}

\bibitem[{{Gheller} \& {Vazza}(2018)}]{Gheller2018}
{Gheller}, C., \& {Vazza}, F. 2018, Monthly Notices of the Royal Astronomical
  Society, 480, 3749, \dodoi{10.1093/mnras/sty2102}

\bibitem[{{Harris} {et~al.}(2020){Harris}, {Millman}, {van der Walt},
  {et~al.}}]{numpy2020}
{Harris}, C.~R., {Millman}, K.~J., {van der Walt}, S.~J., {et~al.} 2020,
  Nature, 585, 357, \dodoi{10.1038/s41586-020-2649-2}

\bibitem[{{Hausen} \& {Robertson}(2020)}]{Hausen2020}
{Hausen}, R., \& {Robertson}, B.~E. 2020, \apjs, 248, 20,
  \dodoi{10.3847/1538-4365/ab8868}

\bibitem[{{Hu} {et~al.}(2018){Hu}, {Shen}, {Albanie}, {Sun}, \&
  {Wu}}]{Hu2018SENet}
{Hu}, J., {Shen}, L., {Albanie}, S., {Sun}, G., \& {Wu}, E. 2018, in
  {Proceedings of the IEEE Conference on Computer Vision and Pattern
  Recognition (CVPR)}, 7132--7141, \dodoi{10.1109/CVPR.2018.00745}

\bibitem[{{Huertas-Company} \& {Lanusse}(2023)}]{HuertasCompany2023}
{Huertas-Company}, M., \& {Lanusse}, F. 2023, \pasa, 40, e001,
  \dodoi{10.1017/pasa.2022.55}

\bibitem[{{Hunter}(2007)}]{matplotlib2007}
{Hunter}, J.~D. 2007, Computing in Science and Engineering, 9, 90,
  \dodoi{10.1109/MCSE.2007.55}

\bibitem[{{Jarrett} {et~al.}(2011){Jarrett}, {Cohen}, {Masci},
  {et~al.}}]{Jarrett2011}
{Jarrett}, T.~H., {Cohen}, M., {Masci}, F., {et~al.} 2011, The Astrophysical
  Journal, 735, 112, \dodoi{10.1088/0004-637X/735/2/112}

\bibitem[{{Lagache} {et~al.}(2005){Lagache}, {Puget}, \& {Dole}}]{Lagache2005}
{Lagache}, G., {Puget}, J.-L., \& {Dole}, H. 2005, Annual Review of Astronomy
  and Astrophysics, 43, 727, \dodoi{10.1146/annurev.astro.43.072103.150606}

\bibitem[{{Lang}(2014)}]{Lang2014}
{Lang}, D. 2014, The Astronomical Journal, 147, 108,
  \dodoi{10.1088/0004-6256/147/5/108}

\bibitem[{{Lauritsen} {et~al.}(2021){Lauritsen}, {Dickinson}, {Bromley},
  {Serjeant}, {Lim}, {Gao}, \& {Wang}}]{Lauritsen2021}
{Lauritsen}, L., {Dickinson}, H., {Bromley}, J., {et~al.} 2021, \mnras, 507,
  1546, \dodoi{10.1093/mnras/stab2195}

\bibitem[{{Law} {et~al.}(2006){Law}, {Mackay}, \& {Baldwin}}]{Law2006}
{Law}, N.~M., {Mackay}, C.~D., \& {Baldwin}, J.~E. 2006, \aap, 446, 739,
  \dodoi{10.1051/0004-6361:20053695}

\bibitem[{{Ledig} {et~al.}(2017){Ledig}, {Theis}, {Husz{\'a}r},
  {et~al.}}]{Ledig2017}
{Ledig}, C., {Theis}, L., {Husz{\'a}r}, F., {et~al.} 2017, in {Proceedings of
  the IEEE Conference on Computer Vision and Pattern Recognition (CVPR)},
  105--114, \dodoi{10.1109/CVPR.2017.19}

\bibitem[{{Liang} {et~al.}(2021){Liang}, {Cao}, {Sun}, {Zhang}, {Van Gool}, \&
  {Timofte}}]{Liang2021SwinIR}
{Liang}, J., {Cao}, J., {Sun}, G., {et~al.} 2021, in {Proceedings of the
  IEEE/CVF International Conference on Computer Vision Workshops (ICCVW)},
  1833--1844, \dodoi{10.1109/ICCVW54120.2021.00210}

\bibitem[{{Lim} {et~al.}(2017){Lim}, {Son}, {Kim}, {Nah}, \&
  {Lee}}]{Lim2017EDSR}
{Lim}, B., {Son}, S., {Kim}, H., {Nah}, S., \& {Lee}, K.~M. 2017, in
  {Proceedings of the IEEE Conference on Computer Vision and Pattern
  Recognition Workshops (CVPRW)}, 1132--1140, \dodoi{10.1109/CVPRW.2017.151}

\bibitem[{{Loshchilov} \& {Hutter}(2019)}]{Loshchilov2019}
{Loshchilov}, I., \& {Hutter}, F. 2019, in {International Conference on
  Learning Representations (ICLR)}

\bibitem[{{Lucy}(1974)}]{Lucy1974}
{Lucy}, L.~B. 1974, \aj, 79, 745, \dodoi{10.1086/111605}

\bibitem[{{Lupton} {et~al.}(1999){Lupton}, {Gunn}, \& {Szalay}}]{Lupton1999}
{Lupton}, R.~H., {Gunn}, J.~E., \& {Szalay}, A.~S. 1999, The Astronomical
  Journal, 118, 1406, \dodoi{10.1086/301004}

\bibitem[{{Mainzer} {et~al.}(2014){Mainzer}, {Bauer}, {Cutri},
  {et~al.}}]{Mainzer2014}
{Mainzer}, A., {Bauer}, J., {Cutri}, R.~M., {et~al.} 2014, The Astrophysical
  Journal, 792, 30, \dodoi{10.1088/0004-637X/792/1/30}

\bibitem[{{Meisner} {et~al.}(2017){Meisner}, {Lang}, \&
  {Schlegel}}]{Meisner2017}
{Meisner}, A.~M., {Lang}, D., \& {Schlegel}, D.~J. 2017, The Astronomical
  Journal, 154, 161, \dodoi{10.3847/1538-3881/aa894e}

\bibitem[{{Moneti} {et~al.}(2022){Moneti}, {McCracken}, {Shuntov}, {Kokorev},
  {Capak}, {Sanders}, {et~al.}}]{Moneti2022}
{Moneti}, A., {McCracken}, H.~J., {Shuntov}, M., {et~al.} 2022, Astronomy \&
  Astrophysics, 658, A126, \dodoi{10.1051/0004-6361/202142361}

\bibitem[{{Narayan} \& {Nityananda}(1986)}]{Narayan1986}
{Narayan}, R., \& {Nityananda}, R. 1986, \araa, 24, 127,
  \dodoi{10.1146/annurev.aa.24.090186.001015}

\bibitem[{{Offringa} \& {Smirnov}(2017)}]{Offringa2017}
{Offringa}, A.~R., \& {Smirnov}, O. 2017, \mnras, 471, 301,
  \dodoi{10.1093/mnras/stx1547}

\bibitem[{{Ramunno} {et~al.}(2025){Ramunno}, {Massa}, {Kinakh}, {Panos},
  {Csillaghy}, \& {Voloshynovskiy}}]{Ramunno2025}
{Ramunno}, F.~P., {Massa}, P., {Kinakh}, V., {et~al.} 2025, \aap, 698, A140,
  \dodoi{10.1051/0004-6361/202453205}

\bibitem[{{Rau} \& {Cornwell}(2011)}]{Rau2011}
{Rau}, U., \& {Cornwell}, T.~J. 2011, \aap, 532, A71,
  \dodoi{10.1051/0004-6361/201117104}

\bibitem[{{Reddy} {et~al.}(2024){Reddy}, {Toomey}, {Parul}, \&
  {Gleyzer}}]{ReddyP2024}
{Reddy}, P., {Toomey}, M.~W., {Parul}, H., \& {Gleyzer}, S. 2024, Machine
  Learning: Science and Technology, 5, 035076, \dodoi{10.1088/2632-2153/ad76f8}

\bibitem[{{Richardson}(1972)}]{Richardson1972}
{Richardson}, W.~H. 1972, Journal of the Optical Society of America, 62, 55,
  \dodoi{10.1364/JOSA.62.000055}

\bibitem[{{Sanders} \& {Mirabel}(1996)}]{Sanders1996}
{Sanders}, D.~B., \& {Mirabel}, I.~F. 1996, \araa, 34, 749,
  \dodoi{10.1146/annurev.astro.34.1.749}

\bibitem[{{Sanders} {et~al.}(2007){Sanders}, {Salvato}, {Aussel},
  {et~al.}}]{Sanders2007}
{Sanders}, D.~B., {Salvato}, M., {Aussel}, H., {et~al.} 2007, The Astrophysical
  Journal Supplement Series, 172, 86, \dodoi{10.1086/517885}

\bibitem[{{Schawinski} {et~al.}(2017){Schawinski}, {Zhang}, {Zhang}, {Fowler},
  \& {Santhanam}}]{Schawinski2017}
{Schawinski}, K., {Zhang}, C., {Zhang}, H., {Fowler}, L., \& {Santhanam}, G.~K.
  2017, Monthly Notices of the Royal Astronomical Society, 467, L110,
  \dodoi{10.1093/mnrasl/slx008}

\bibitem[{{Scoville} {et~al.}(2007){Scoville}, {Aussel}, {Brusa},
  {et~al.}}]{Scoville2007}
{Scoville}, N., {Aussel}, H., {Brusa}, M., {et~al.} 2007, The Astrophysical
  Journal Supplement Series, 172, 1, \dodoi{10.1086/516585}

\bibitem[{{Shi} {et~al.}(2016){Shi}, {Caballero}, {Husz{\'a}r},
  {et~al.}}]{Shi2016}
{Shi}, W., {Caballero}, J., {Husz{\'a}r}, F., {et~al.} 2016, in {Proceedings of
  the IEEE Conference on Computer Vision and Pattern Recognition (CVPR)},
  1874--1883, \dodoi{10.1109/CVPR.2016.207}

\bibitem[{{Song} {et~al.}(2024){Song}, {Ma}, {Sun}, {Zhao}, \&
  {Lin}}]{Song2024}
{Song}, W., {Ma}, Y., {Sun}, H., {Zhao}, X., \& {Lin}, G. 2024, \aap, 686,
  A272, \dodoi{10.1051/0004-6361/202349100}

\bibitem[{{Sukurdeep} {et~al.}(2025){Sukurdeep}, {Budav{\'a}ri}, {Connolly}, \&
  {Navarro}}]{Sukurdeep2025}
{Sukurdeep}, Y., {Budav{\'a}ri}, T., {Connolly}, A.~J., \& {Navarro}, F. 2025,
  \aj, 170, 233, \dodoi{10.3847/1538-3881/adfb72}

\bibitem[{{Sweere} {et~al.}(2022){Sweere}, {Valtchanov}, {Lieu}, {Vojtekova},
  {Verdugo}, {Santos-Lleo}, {Pacaud}, {Briassouli}, \& {C{\'a}mpora
  P{\'e}rez}}]{Sweere2022}
{Sweere}, S.~F., {Valtchanov}, I., {Lieu}, M., {et~al.} 2022, \mnras, 517,
  4054, \dodoi{10.1093/mnras/stac2437}

\bibitem[{{Traina} {et~al.}(2024){Traina}, {Gruppioni}, {Delvecchio},
  {et~al.}}]{Traina2024}
{Traina}, A., {Gruppioni}, C., {Delvecchio}, I., {et~al.} 2024, \aap, 681,
  A118, \dodoi{10.1051/0004-6361/202347048}

\bibitem[{{unWISE Team}(2021)}]{unWISEimages2021}
{unWISE Team}. 2021, {unWISE Images},  IPAC, \dodoi{10.26131/IRSA524}

\bibitem[{{van der Walt} {et~al.}(2014){van der Walt}, {Sch{\"o}nberger},
  {Nunez-Iglesias}, {Boulogne}, {Warner}, {Yager}, {Gouillart}, {Yu}, \& {the
  scikit-image contributors}}]{scikitimage2014}
{van der Walt}, S., {Sch{\"o}nberger}, J.~L., {Nunez-Iglesias}, J., {et~al.}
  2014, PeerJ, 2, e453, \dodoi{10.7717/peerj.453}

\bibitem[{{Wang} {et~al.}(2019){Wang}, {Yu}, {Wu}, {Gu}, {Liu}, {Dong}, {Qiao},
  \& {Loy}}]{Wang2018ESRGAN}
{Wang}, X., {Yu}, K., {Wu}, S., {et~al.} 2019, in {Computer Vision -- ECCV 2018
  Workshops}, 63--79, \dodoi{10.1007/978-3-030-11021-5\_5}

\bibitem[{{Wang} {et~al.}(2004){Wang}, {Bovik}, {Sheikh}, \&
  {Simoncelli}}]{Wang2004}
{Wang}, Z., {Bovik}, A.~C., {Sheikh}, H.~R., \& {Simoncelli}, E.~P. 2004, IEEE
  Transactions on Image Processing, 13, 600, \dodoi{10.1109/TIP.2003.819861}

\bibitem[{{Weaver} {et~al.}(2022){Weaver}, {Kauffmann}, {Ilbert}, {McCracken},
  {Moneti}, {Toft}, {Brammer}, {Shuntov}, {Davidzon}, {Hsieh}, {Laigle},
  {Mobasher}, {Sanders}, {et~al.}}]{Weaver2022}
{Weaver}, J.~R., {Kauffmann}, O.~B., {Ilbert}, O., {et~al.} 2022, \apjs, 258,
  11, \dodoi{10.3847/1538-4365/ac3078}

\bibitem[{{Werner} {et~al.}(2004){Werner}, {Roellig}, {Low},
  {et~al.}}]{Werner2004}
{Werner}, M.~W., {Roellig}, T.~L., {Low}, F.~J., {et~al.} 2004, The
  Astrophysical Journal Supplement Series, 154, 1, \dodoi{10.1086/422992}

\bibitem[{{Wright} {et~al.}(2010){Wright}, {Eisenhardt}, {Mainzer},
  {et~al.}}]{Wright2010}
{Wright}, E.~L., {Eisenhardt}, P.~R.~M., {Mainzer}, A.~K., {et~al.} 2010, The
  Astronomical Journal, 140, 1868, \dodoi{10.1088/0004-6256/140/6/1868}

\bibitem[{{Zaroubi} {et~al.}(1995){Zaroubi}, {Hoffman}, {Fisher}, \&
  {Lahav}}]{Zaroubi1995}
{Zaroubi}, S., {Hoffman}, Y., {Fisher}, K.~B., \& {Lahav}, O. 1995, \apj, 449,
  446, \dodoi{10.1086/176070}

\bibitem[{{Zhang} {et~al.}(2018){Zhang}, {Li}, {Li}, {Wang}, {Zhong}, \&
  {Fu}}]{Zhang2018RCAN}
{Zhang}, Y., {Li}, K., {Li}, K., {et~al.} 2018, in {Proceedings of the European
  Conference on Computer Vision (ECCV)}, 294--310,
  \dodoi{10.1007/978-3-030-01234-2\_18}

\bibitem[{{Zhao} {et~al.}(2017){Zhao}, {Gallo}, {Frosio}, \&
  {Kautz}}]{Zhao2017}
{Zhao}, H., {Gallo}, O., {Frosio}, I., \& {Kautz}, J. 2017, IEEE Transactions
  on Computational Imaging, 3, 47, \dodoi{10.1109/TCI.2016.2644865}

\end{thebibliography}

\end{document}